\documentclass[12pt]{article}
\usepackage{amsfonts}
\usepackage{amssymb}
\usepackage{graphicx}
\usepackage{amsmath}
\usepackage{makeidx}
\usepackage{indentfirst}
\usepackage[T1]{fontenc}
\usepackage[utf8]{inputenc}
\usepackage{authblk}

\newcounter{resultnum}[section]
\newcounter{conclusionnum}[section]
\newcounter{conditionnum}[section]
\newcounter{conjecturenum}[section]
\newcounter{examplenum}[section]
\newcounter{exercisenum}[section]
\newcounter{lemmanum}[section]
\newcounter{notationnum}[section]
\newcounter{theoremnum}[section]
\newcounter{definitionnum}[section]
\newcounter{corollarynum}[section]
\newcounter{remarknum}[section]
\newcounter{propositionnum}[section]
\newcounter{acknowledgementnum}[section]
\newcounter{algorithmnum}[section]
\newcounter{axiomnum}[section]
\newcounter{casenum}[section]
\newcounter{claimnum}[section]
\newcounter{summarynum}[section]
\newcounter{problemnum}[section]
\begin{document}

\title{Anamorphic Quasiperiodic Universes in Modified and Einstein Gravity
with Loop Quantum Gravity Corrections}
\date{July 17, 2017}
\author{${}$ \\
Marcelo M. Amaral\\
{\small \textit{Quantum Gravity Research; 101 S. Topanga Canyon Blvd \#
1159. Topanga, CA 90290, USA}} \\
{\small \textit{emails: marcelo@quantumgravityresearch.org }}\\
${}$ \\
Raymond Aschheim\\
{\small \textit{Quantum Gravity Research; 101 S. Topanga Canyon Blvd \#
1159. Topanga, CA 90290, USA}} \\
{\small \textit{email: raymond@quantumgravityresearch.org }}\\
${}$ \\
Lauren\c tiu Bubuianu\\
{\small \textit{TVR Ia\c si, \ 33 Lasc\v ar Catargi street; and University Apollonia, 2 Muzicii street,  700107 Ia\c si,
Romania }} \\
{\small \textit{email: laurentiu.bubuianu@tvr.ro }}\\
${}$ \\
Klee Irwin \\
{\small \textit{Quantum Gravity Research; 101 S. Topanga Canyon Blvd \#
1159. Topanga, CA 90290, USA}} \\
{\small \textit{email: klee@quantumgravityresearch.org }}\\
${}$ \\
Sergiu I. Vacaru\thanks{ \textit{for postal correspondence with this author use:\ } 67 Lloyd Street South, Manchester, M14 7LF, the UK;  \\ the affiliation and e-mails are for a visiting position in the USA}\\
{\small \textit{ Physics Department, California State University Fresno, Fresno, CA 93740, USA}} \\

{\small and } {\small \textit{University "Al. I. Cuza" Ia\c si, Project IDEI, Ia\c si, Romania }} \\
{\small \textit{email: sergiu.vacaru@gmail.com and dougs@csufrenso.edu}} \\
${}$ \\
Daniel Woolridge \\
{\small \textit{Quantum Gravity Research; 101 S. Topanga Canyon Blvd \#
1159. Topanga, CA 90290, USA}} \\
{\small \textit{email: dan@quantumgravityresearch.org }}
}

\maketitle

\begin{abstract}
The goal of this work is to elaborate on new geometric methods of
constructing exact and parametric quasiperiodic solutions for anamorphic
cosmology models in modified gravity theories, MGTs, and general relativity,
GR. There exist previously studied generic off-diagonal and diagonalizable
cosmological metrics encoding gravitational and matter fields with
quasicrystal like structures, QC, and holonomy corrections from loop quantum
gravity, LQG. We apply the anholonomic frame deformation method, AFDM, in
order to decouple the (modified) gravitational and matter field equations in
general form. This allows us to find integral varieties of cosmological
solutions determined by generating functions, effective sources, integration
functions and constants. The coefficients of metrics and connections for
such cosmological configurations depend, in general, on all spacetime
coordinates and can be chosen to generate observable (quasi)-periodic/
aperiodic/ fractal / stochastic / (super) cluster / filament / polymer like
(continuous, stochastic, fractal and/or discrete structures) in MGTs and/or
GR. In this work, we study new classes of solutions for anamorphic cosmology
with LQG holonomy corrections. Such solutions are characterized by nonlinear
symmetries of generating functions for generic off--diagonal cosmological
metrics and generalized connections, with possible nonholonomic constraints
to Levi--Civita configurations and diagonalizable metrics depending only on
a time like coordinate. We argue that anamorphic quasiperiodic cosmological
models integrate the concept of quantum discrete spacetime, with certain
gravitational QC-like vacuum and nonvacuum structures. And, that of a
contracting universe that homogenizes, isotropizes and flattens without
introducing initial conditions or multiverse problems.

\vskip5pt

\textbf{Keywords:}\ Mathematical cosmology; geometry of nonholonomic
spacetimes; post modern inflation paradigm; ekpyrotic universes; modified
gravity theories; loop quantum gravity and cosmology; quasiperiodic
cosmological structures.

\vskip10pt


\end{abstract}

\tableofcontents



\section{ Introduction and Motivation}

It is thought that near the Planck limit any quantum gravity theory is
characterized by discrete degrees of freedom, respective of quantum minimal
length and quantum symmetries, and anisotropic and inhomogeneous fluctuating
/ random configurations. On the other hand, observations show that the
accelerating Universe is flat, smooth and scale free at large-scale
distances when the spectrum of primordial curvature perturbations is nearly
scale-invariant, adiabatic and Gaussian \cite{sht1a,sht2a}. We cite papers
\cite{sht3a,sht4a,sht5a,ijj,guthah01,kallosh1,mukh01,starob01} for recent
reviews, discussions, critique and new results on postmodern inflation
scenarios developed and advocated by prominent theorists in relation to the
Planck 2013 and Planck 2015 cosmological data \cite%
{planck13a16,planck13a22,planck15a13,planck15a14,planck15a20,planck15a26,planck15a31}%
. Here we note that for meta-galactic and galactic distances, the Planck
2015 and WMAP, ACT and SPT teams' observation and theoretical results%
\footnote{%
consistency and implications for inflationary, ekpyrotic and anamorphic
bouncing cosmologies, and other type cosmological models, are discussed in
Refs. \cite{sht4a,planck15a31}} on spacetime anisotropy and
topology, dark energy, and constraints on inflation and accelerating
cosmology parameters. Such works conclude on the existence of mixed
aperiodic and quasiperiodic structures (for gravitational, dark matter and
standard matter) described as net-works for the first group- and (super)
cluster-scale, strong gravitational lensing / light filaments / polymer and
quasicrystal, QC, like configurations.

In our partner works \cite{partn1,partn2,partn3}, we proved that
Starobinsky-like inflation \cite{starob} and various dark energy, DE, and
dark matter, DM, effects in a Universe with quasiperiodic (super) cluster
and filament configurations can be determined by a nontrivial QC spacetime
structure. We cite see Refs. \cite%
{steinh1,steinh2,steinh3,steinh4,emmr,subr,penrose1,penrose2,achim,barkan,qcalloys1,qcalloys2,qcalloys3,qcalloys4,qcalloys5,crystalinks}
for important works and references on the physics and mathematics of QCs in
condensed matter physics but also with possible connections to cosmology.
Various F--modified (for instance, $F(R)=R+\alpha R^{2}$) cosmological models%
\footnote{%
in more general contexts, one considers various modified gravity theories, $%
F(R,\mathcal{T}),F(T),...$ determined by functionals on Ricci scalars,
energy-momentum and/or torsion tensors etc.; in various papers, such
functionals are denoted also as $f(R)$, $f(T)$,...} can be with
singularities and encode inhomogeneous and locally anisotropic properties.
For reviews on modified gravity theories, MGTs, readers may consider \cite%
{capoz,nojod1,clifton,gorbunov,linde2,bambaodin,sami,mimet2,odints2,guend1,guend2,cosmv2,eegrg,eeejpp,cosmv3,stavr1,stavr2}%
. In papers \cite{amoros1,amoros2}, a detailed analytical and numerical
study of possible holonomy corrections from LQG to $f(R)$ gravity was
performed. It was shown that, as a result of such quantum corrections (and
various generic off-diagonal, nonholonomic and/or QC contributions
investigated in \cite{partn1,partn2,partn3,cosmv2,eegrg,eeejpp,cosmv3}) the
dynamics may change substantially and, for certain well defined conditions,
one obtains better predictions for the inflationary phase as compared with
current observations. Various approaches to LQG and spin network theories,
see also constructions on loop quantum cosmology, LQC, are reviewed in Refs.
\cite{rov,thiem1,asht,asht0,boj}. In the past, certain criticism against LQG
(see, for instance, \cite{nicolai}) was motivated in the bulk by arguments
that the mathematical formalism is not that which is familiar for the
particle physicists working with perturbation theory, Fock spaces,
background fields etc; see reply and discussion in \cite{thiem}.

The main objectives of this work are to study how quasiperiodic and/or
aperiodic QC like structures with possible holonomy LQG corrections modify
inflation and acceleration cosmology scenarios in MGTs and GR; to analyse if
such effects can be modeled in the framework of the Einstein gravity theory;
and to show how such generic off--diagonal cosmological solutions can be
constructed and treated in anamorphic cosmology. The extensions of
cosmological models to spacetimes with nontrivial quasiperiodic/ aperiodic
and general anistoropic structures is not a trivial task. It is necessary to
elaborate on new classes of exact and/or parametric solutions of
gravitational and matter field cosmological equations which, in general,
depend on all spacetime variables via generating and integration functions
with mixed smooth and discrete degree of freedom and anisotropically
polarized physical constants. We emphasize that it is not possible to
describe, for instance, growth of any QC structure and compute certain
cosmological effects determined by non-perturbative and nonlinear
gravitational interactions if we restrict our models to only diagonal
homogeneous and isotropic metrics like the Friedmann--Lama\^{\i}%
tre-Roberstson-Worker, FLRW, one and possible generalizations with Lie
group/algebroid symmetries \cite{cosmv2,eegrg,eeejpp,cosmv3}. In such cases,
the cosmological solutions are determined by some integration and/or
structure group constants, and depend only on a time like coordinate. We can
not describe in a realistic form quasiperiodic / aperiodic spacetime
structures, and their evolution, using only time-depending functions and
FLRW metrics. In order to formulate and develop an unified geometric
approach for all observational data on (super) cluster and extra long
cosmological distances, we have to work with "non-diagonalizable" metrics%
\footnote{%
which can not be diagonalized by coordinate transforms, in a local or
infinite spacetime region} and generalized connections, and apply new
numeric and analytic methods for constructing more general classes of
solutions in MGTs and cosmology models with quasiperiodic structure,
inhomogeneities and local anisotropies. The new classes of cosmological
solutions incorporate generating functions and integration functions, with
various integration constants and parameters, which allow more opportunities
to compare with experimental data. Even some subclasses of solutions can be
parameterized by effective diagonal metrics\footnote{%
for certain limits with small off--diagonal corrections and/or
nonholonomically constrained configurations, for instance, incorporating
anomorphic smoothing phases}, the diagonal coefficients contain various
physical data of nonlinear classical and quantum interactions encoded via
generating functions and effective sources.

In contrast to the general purpose of unification of physical interactions
and development of fundamental and geometric principles of quantization (for
instance, in string theory and deformation quantization), the approaches
based on LQG and spin networks were performed originally just as theories of
quantum gravity combining the general relativity (GR) and quantum mechanics.
The main principle was to provide a non--perturbative formulation when the
background independence (the key feature of Einstein's theory) is preserved.
At the present time, LQG is supposed to have a clear conceptual and logical
setup following from physical considerations and supported by a rigorous
mathematical formulation. In this work, we study a toy cosmological model
with LQG contributions, whilst keeping in mind that such constructions will
be expanded on for spin network models and further generalizations to QC
configurations. Here, in addition to the references presented above, we cite
some fundamental works \cite{asht1,asht2,barb1,barb2,immirzi,vlqgdq} on LQG
for also considering developments in loop quantum cosmology and possible
extensions, for example, to deformation quantization. We emphasize that we
analyze examples with a special class of holonomy corrections from LQG in
order to prove that possible quantum modifications do not affect the main
results on anamorphic cosmological models with QC structure.

With respect to our toy LQC model, we also note that we restrict our study
to quantum gravity quasiperiodic effects in anamorphic cosmology by
considering a special class of holonomy corrections from LQG in order to
distinguish possible non-perturbative and background independent
modifications. In this approach, quantization can be performed in certain
forms preserving the Lorentz local invariance in the continuous limit. Here
we note that if the quantization formalism is developed on (co)tangent
bundles, one gets quantum corrections and respective cosmological terms
violating this local symmetry \cite{vcosmsol1}. In a more general context,
such an approach involves reformulation of the LQG in nonholonomic variables
with double 2+2 and 3+1 fibrations considered in \cite{vlqgdq,vdoublefibr}.
Details on the so--called ADM, i.e. Arnowitt--Deser--Misner, formalism in GR
can be found, for instance, in \cite{mtw,rov,thiem1,asht}. In order to
construct new classes of cosmological solutions, we shall apply the
anholonomic frame deformation method, AFDM (see details and examples for
accelerating cosmology and DE and DM physics in \cite%
{gvvafdm,vanp,vrajssrf,cosmv2,cosmv3}).

The paper is organized as follows: In section \ref{sec2nhalqc}, we outline
the most important formulas on nonholonomic variables, frame, linear and
nonlinear connection deformations used for constructing (in general) generic
off--diagonal cosmological solutions depending on all spacetime coordinates.
It is shown how using such constructions we can decouple the gravitational
and matter field equations in accelerating cosmology if the Einstein gravity
and various $f(R)$ modifications, with LQG corrections. In nonholonomic
variables we formulate the criteria for anamorphic cosmological phases and
analyze possible small parametric deformations in terms of quasi-FLRW
metrics for nonholonomic Friedmann equations.

Section \ref{sectodancosm} is devoted to the study of geometric properties
of new classes of generic off--diagonal cosmological solutions modeling QC
like structures in MGTs with LQG sources. In this section the conditions on
generating and integration functions and integration constants when such
configurations encode quasiperiodic/ aperiodic structure of possible
different origin (induced by F-modifications, gravitational like
polarization of mass like constants, anamorphic phases with effective
polarization of the cosmological constant, and LQC sources) are formulated.
Four such classes of solutions are constructed in explicit form and the
criteria for anamorphic QC phases are formulated. Here, we also provide
solutions for nonlinear superpositions resulting in hierarchies with new
anamorphic QC like cosmological solutions.

In section \ref{epsilonpar}, we consider small parametric decompositions for
quasi-FLRW metrics encoding QC like structures. It is proven that in such
cases the cosmological solutions with gravitationally polarized cosmological
constants and the criteria for anamorphic phases can be written in certain
forms similar to homogeneous cosmological configurations. In such cases, QC
and LQG modified Friedmann equations can be derived in explicit form.

We discuss the results in section \ref{secconcl}. Appendix \ref{appendixsect} provides a summary on geometric methods for constructing off-diagonal and diagonal cosmological solutions.

\section{Nonholonomic Variables and Anamorphic Cosmology}

\label{sec2nhalqc}To be able to construct, in explicit form, exact and
parametric quasiperiodic cosmological solutions in MGTs with quantum
corrections we have to re--write the fundamental gravitational and matter
field equations in such nonholonomic variables when a decoupling and general
integration of corresponding systems of nonlinear partial differential
equations, PDEs, are possible. Readers are referred to \cite%
{vcosmsol1,vdoublefibr,gvvafdm,vanp,vrajssrf,cosmv2,cosmv3} for details on
the geometry and applications of the AFDM as a method of constructing exact
solutions in gravity and Ricci flow theories. In this section, we show how
such nonholonomic variables can be introduced in MGT and GR theory and
formulate a geometric approach to anamorphic cosmology \cite%
{sht1a,sht2a,sht3a,sht4a,sht5a,ijj}. The constructions will be used in the
next section for decoupling the fundamental cosmological PDEs with matter
field sources and LQG corrections parameterized as in Refs. \cite%
{zhang1,zhang2,amoros1,amoros2,vlqgdq,bamba,amoros}.

\subsection{N-adapted frames and connection deformations in MGTs}

We presume that the metric properties of a four dimensional, 4d,
cosmological spacetime manifold $\mathbf{V}$ are defined by a metric $%
\mathbf{g}$ of pseudo--Riemannian signature $(+++-)$ which can be
parameterized as a distinguished metric, d--metric,
\begin{eqnarray}
\mathbf{g} &=&\mathbf{g}_{\alpha \beta }(u)\mathbf{e}^{\alpha }\otimes
\mathbf{e}^{\beta }=g_{i}(x^{k})dx^{i}\otimes dx^{i}+g_{a}(x^{k},y^{b})%
\mathbf{e}^{a}\otimes \mathbf{e}^{b}  \label{odans} \\
&=&\mathbf{g}_{\alpha ^{\prime }\beta ^{\prime }}(u)\mathbf{e}^{\alpha
^{\prime }}\otimes \mathbf{e}^{\beta ^{\prime }},\mbox{ for }\mathbf{g}%
_{\alpha ^{\prime }\beta ^{\prime }}(u)=\mathbf{g}_{\alpha \beta }\mathbf{e}%
_{\ \alpha ^{\prime }}^{\alpha }\mathbf{e}_{\ \beta ^{\prime }}^{\beta }.
\notag
\end{eqnarray}%
In these formulas, we use N--adapted frames, \ $\mathbf{e}_{\alpha }=(%
\mathbf{e}_{i},e_{a}),$ and dual frames, $\mathbf{e}^{\alpha }=(x^{i},%
\mathbf{e}^{a}),$
\begin{eqnarray}
\mathbf{e}_{i} &=&\partial /\partial x^{i}-N_{i}^{a}(u)\partial /\partial
y^{a},e_{a}=\partial _{a}=\partial /\partial y^{a},  \label{nder} \\
e^{i} &=&dx^{i},\mathbf{e}^{a}=dy^{a}+N_{i}^{a}(u^{\gamma })dx^{i}%
\mbox{ and
}\mathbf{e}^{\alpha }=\mathbf{e}_{\ \alpha ^{\prime }}^{\alpha
}(u)du^{\alpha ^{\prime }}.  \notag
\end{eqnarray}%
The local coordinates on $\mathbf{V}$ are labeled $u^{\gamma }=(x^{k},y^{c}),
$ or $u=(x,y),$ when indices run corresponding values $i,j,k,...=1,2$ and $%
a,b,c,...=3,4$ (for nonholonomic 2+2 splitting, for $u^{4}=y^{4}=t$ being a
time like coordinate and $u^{\grave{\imath}}=(x^{i},y^{3})$ considered as
spacelike coordinates endowed with indices $\grave{\imath},\grave{j},\grave{k%
},...=1,2,3.$ We note that a local basis\footnote{%
in literature, one uses equivalent terms like frame, tetrad, vierbein systems%
} $\mathbf{e}_{\alpha }$ is nonholonomic (equivalently, non-integrable, or
anholonomic) if the commutators
\begin{equation}
\mathbf{e}_{[\alpha }\mathbf{e}_{\beta ]}:=\mathbf{e}_{\alpha }\mathbf{e}%
_{\beta }-\mathbf{e}_{\beta }\mathbf{e}_{\alpha }=C_{\alpha \beta }^{\gamma
}(u)\mathbf{e}_{\gamma }  \label{anhr}
\end{equation}%
contain nontrivial anholonomy coefficients $C_{\alpha \beta }^{\gamma
}=\{C_{ia}^{b}=\partial _{a}N_{i}^{b},C_{ji}^{a}=\mathbf{e}_{j}N_{i}^{a}-%
\mathbf{e}_{i}N_{j}^{a}\}.$

A value $\mathbf{N}=\{N_{i}^{a}\}=N_{i}^{a}\frac{\partial }{\partial y^{a}}%
\otimes dx^{i}$ determined by frame coefficients in (\ref{nder}) defines a
nonlinear connection, N-connection, structure as an N--adapted decomposition
of the tangent bundle
\begin{equation}
T\mathbf{V}=hT\mathbf{V}\oplus vT\mathbf{V}  \label{ncon}
\end{equation}%
into conventional horizontal, h, and vertical, v, subspaces. On a 4-d
metric--affine manifold $\mathbf{V,}$ this states an equivalent fibred
structure with nonholonomic 2+2 spacetime decomposition (splitting). In
particular, such a h-v-splitting states a double, \ $h$ and $v,$ diadic
frame structure on any (pseudo) Riemannian spacetime. We shall use boldface
symbols for geometric/ physical objects on a spacetime manifold $\mathbf{V}$
endowed with geometric objects $\left( \mathbf{g,N,D}\right) $. The values $%
\mathbf{D}$ is a distinguished connection, d--connection, $\mathbf{D}%
=(hD,vD) $ defined as a linear connection, i.e. a metric--affine one,
preserving the $N$--connection splitting (\ref{ncon}) under parallel
transports. We denote by $\mathcal{T}=\{\mathbf{T}_{\beta \gamma }^{\alpha
}\}$ the torsion of $\mathbf{D,}$ which can be computed in standard form,
see geometric preliminaries in \cite%
{vlqgdq,vcosmsol1,vdoublefibr,gvvafdm,vanp,vrajssrf,cosmv2,cosmv3}.

On a nonholonomic spacetime manifold $\mathbf{V}$, we can work equivalently
with two linear connections defined by the same metric structure $\mathbf{g}$%
:
\begin{equation}
(\mathbf{g,N})\rightarrow \left\{
\begin{array}{cc}
\mathbf{\nabla :} & \mathbf{\nabla g}=0;\ ^{\nabla }\mathcal{T}=0,%
\mbox{\
for  the Levi-Civita,  LC, -connection } \\
\widehat{\mathbf{D}}: & \widehat{\mathbf{D}}\mathbf{g}=0;\ h\widehat{%
\mathcal{T}}=0,v\widehat{\mathcal{T}}=0,hv\widehat{\mathcal{T}}\neq 0,%
\mbox{
for the canonical d--connection  }.%
\end{array}%
\right.  \label{twocon}
\end{equation}%
As a result, it is possible to formulate equivalent models of
pseudo-Riemannian geometry and/or Riemann--Cartan geometry with
nonholonomically induced torsion\footnote{\label{notericci}It should be
emphasized that the canonical distortion relation $\widehat{\mathbf{D}}%
=\nabla +\widehat{\mathbf{Z}}$, where the distortion distinguished tensor,
d-tensor, $\widehat{\mathbf{Z}}=\{\widehat{\mathbf{Z}}_{\ \beta \gamma
}^{\alpha }[\widehat{\mathbf{T}}_{\ \beta \gamma }^{\alpha }]\},$ is an
algebraic combination of the coefficients of the corresponding torsion
d-tensor $\widehat{\mathcal{T}}=\{\widehat{\mathbf{T}}_{\ \beta \gamma
}^{\alpha }\}$ of $\widehat{\mathbf{D}}.$ The curvature tensors of both
linear connections are computed in standard forms, $\widehat{\mathcal{R}}=\{%
\widehat{\mathbf{R}}_{\ \beta \gamma \delta }^{\alpha }\}$ and $\ ^{\nabla }%
\mathcal{R}=\{R_{\ \beta \gamma \delta }^{\alpha }\}$ (respectively, for $%
\widehat{\mathbf{D}}$ and $\nabla ).$ This allows us to introduce the
corresponding Ricci tensors, $\ \widehat{\mathcal{R}}ic=\{\widehat{\mathbf{R}%
}_{\ \beta \gamma }:=\widehat{\mathbf{R}}_{\ \alpha \beta \gamma }^{\gamma
}\}$ and $Ric=\{R_{\ \beta \gamma }:=R_{\ \alpha \beta \gamma }^{\gamma }\}$%
. The value $\widehat{\mathcal{R}}ic$ is characterized by $h$-$v$ N-adapted
coefficients, $\widehat{\mathbf{R}}_{\alpha \beta }=\{\widehat{R}_{ij}:=%
\widehat{R}_{\ ijk}^{k},\ \widehat{R}_{ia}:=-\widehat{R}_{\ ika}^{k},\
\widehat{R}_{ai}:=\widehat{R}_{\ aib}^{b},\ \widehat{R}_{ab}:=\widehat{R}_{\
abc}^{c}\}.$ There are also two different scalar curvatures, $\ R:=\mathbf{g}%
^{\alpha \beta }R_{\alpha \beta }$ and $\ \widehat{\mathbf{R}}:=\mathbf{g}%
^{\alpha \beta }\widehat{\mathbf{R}}_{\alpha \beta }=g^{ij}\widehat{R}%
_{ij}+g^{ab}\widehat{R}_{ab}$. We can also consider additional constraints
resulting in zero values for the canonical d-torsion, $\widehat{\mathcal{T}}%
=0,$ considering some limits $\widehat{\mathbf{D}}_{\mid \widehat{\mathcal{T}%
}\rightarrow 0}=\mathbf{\nabla .}$}, see a summary of most important
formulas in Appendix \ref{appendixsect}.

\subsection{Anamorphic cosmology in nonholonomic variables}

Based on invariant criteria, authors \cite{sht1a,sht2a,sht3a,sht4a,sht5a,ijj}
attempted to develop a complete scenario explaining the smoothness and
flatness of the universe on large scales with a smoothing phase that acts
like a contracting universe. In this section, we develop a model of
anamorphic cosmology in the framework of MGTs with quasiperiodic / aperiodic
structures and LQC--corrections. The approach relies on having time-varying
masses for particles and certain Weyl-invariant values that define certain
aspects of contracting and/or expanding cosmological backgrounds. For
off-diagonal cosmological models with nontrivial vacuum structures, the
variation of masses and physical constants have a natural explanation via
gravitational polarization functions \cite%
{partn1,partn2,partn3,cosmv2,eegrg,eeejpp,cosmv3}. Let us denote such
variations of a particle mass $m\rightarrow \check{m}(x^{i},t)\simeq \check{m%
}(t)$ and of Planck mass $M_{P}$ $\rightarrow \check{M}_{P}(x^{i},t)\simeq
\check{M}_{P}(t),$ which depends on the type of generating functions we
consider. The actions for particle motion and modified gravity are written
respectively as
\begin{eqnarray}
~^{p}\mathcal{S} &\mathcal{=}&\int \frac{\check{m}}{\check{M}_{P}}ds%
\mbox{
and }  \label{massact} \\
\mathcal{S} &=&\int d^{4}u\sqrt{|\mathbf{g}|}[\mathbf{F}(\widehat{\mathbf{R}}%
)+~^{m}\mathcal{L}(\phi )]  \label{modanamact} \\
&=&\int d^{4}u\sqrt{|\mathbf{g}|}[\frac{1}{2}\check{M}_{P}^{2}(\phi )%
\widehat{\mathbf{R}}-\frac{1}{2}\kappa (\phi )\mathbf{g}^{\alpha \beta }(%
\mathbf{e}_{\alpha }\phi )(\mathbf{e}_{\beta }\phi )-~^{J}V(\phi )+~^{m}%
\mathcal{L}(\phi )],  \label{modanamact1}
\end{eqnarray}%
where $\check{M}_{P}^{2}(\phi ):=M_{Pl}^{0}\sqrt{f(\phi )}$ is positive
definite (we can work in a system of coordinates when $M_{Pl}^{0}=1$). Above
actions are written for a d--metric $\mathbf{g}_{\alpha \beta }$ (\ref{odans}%
), $\kappa (\phi )$ is the nonlinear kinetic coupling function and $\widehat{%
\mathbf{R}}$ is the scalar curvature of $\widehat{\mathbf{D}}.$ In our
works, we use left labels in order to denote, for instance, that \ $~^{m}%
\mathcal{L}$ is for matter fields (for this label, $m$ is from "mass") and $%
~^{J}V$ is for the Jordan frame representation.\footnote{%
If in (\ref{modanamact}) and (\ref{modanamact1}) $\mathbf{F}(\widehat{%
\mathbf{R}})=\widehat{\mathbf{R}}^{2},$ $~^{J}V(\phi )=-\Lambda $ and $~^{m}%
\mathcal{L}=0,$ we obtain a quadratic action for nonholonomic MGTs studied
in \cite{partn1,partn2,partn3,vanp,vrajssrf}, when $\mathcal{S}=\int d^{4}u%
\sqrt{|\mathbf{g}|}[\widehat{\mathbf{R}}^{2}+~^{m}\mathcal{L}].$ The
equivalence of such actions to nonholonomic deformations of the Einstein
gravity with scalar field sources can be derived from the invariance (both
for $\nabla $ and $\widehat{\mathbf{D}}$) under global dilatation symmetry
with a constant $\sigma ,$ $g_{\mu \nu }\rightarrow e^{-2\sigma }g_{\mu \nu
},\phi \rightarrow e^{2\sigma }\widetilde{\phi }.$ We can re-define the
physical values from the Jordan to the Einstein, E, frame using $\phi =\sqrt{%
3/2}\ln |2\widetilde{\phi }|,$ when $\ ^{E}\mathcal{S}=\int d^{4}u\sqrt{|g|}%
\left( \frac{1}{2}\widehat{\mathbf{R}}-\frac{1}{2}\mathbf{e}_{\mu }\phi \
\mathbf{e}^{\mu }\phi -2\Lambda \right) $. The field equations derived from $%
\ ^{E}\mathcal{S}$ are
\begin{equation*}
\widehat{\mathbf{R}}_{\mu \nu }-\mathbf{e}_{\mu }\phi \ \mathbf{e}_{\nu
}\phi -2\Lambda \mathbf{g}_{\mu \nu }=0\mbox{ and }\widehat{\mathbf{D}}%
^{2}\phi =0.
\end{equation*}%
To find explicit solutions we can consider $~\mathbf{\Upsilon }_{\mu \nu
}\sim diag[0,0,~\mathbf{\Upsilon ,}~\mathbf{\Upsilon }],$ where $~\mathbf{%
\Upsilon (}t\mathbf{)}$ will be determined by scalar fields in anamorphic QC
phase and possible holonomic corrections to the Hubble constant. We obtain
the Einstein gravity theory if $\mathbf{F}(\widehat{\mathbf{R}})=R$ for $%
\widehat{\mathbf{D}}_{\mid \widehat{\mathcal{T}}\rightarrow 0}=\mathbf{%
\nabla .}$ For simplicity, we can consider matter actions $~^{m}\mathcal{S}%
=\int d^{4}u\sqrt{|\mathbf{g}|}~^{m}\mathcal{L}$ for matter field Lagrange
densities $~^{m}\mathcal{L}$ depending only on coefficients of a metric
field and do not depend on their derivatives when
\begin{equation*}
\ ^{m}\mathbf{T}_{\alpha \beta }:=-\frac{2}{\sqrt{|\mathbf{g}_{\mu \nu }|}}%
\frac{\delta (\sqrt{|\mathbf{g}_{\mu \nu }|}\ \ ^{m}\mathcal{L})}{\delta
\mathbf{g}^{\alpha \beta }}=\ ^{m}\mathcal{L}\mathbf{g}^{\alpha \beta }+2%
\frac{\delta (\ ^{m}\mathcal{L})}{\delta \mathbf{g}_{\alpha \beta }}.
\end{equation*}%
} Here we note that $\mathbf{F}(\widehat{\mathbf{R}})=\mathbf{F}[\widehat{%
\mathbf{R}}(\mathbf{g},\widehat{\mathbf{D}},\phi )]$ is also a functional of
the scalar field $\phi $ but we use simplified notations using the
assumption that $\widehat{\mathbf{R}}(\mathbf{g,}\widehat{\mathbf{D}})$ are
related to $\phi $ by a source term for modified Einstein equations with
such a nonlinear scalar field.

The gravitational field equations for MGT with functional $\mathbf{F}(%
\widehat{\mathbf{R}})$ in (\ref{modanamact}) can be derived by a N--adapted
variational calculus, see details in \cite%
{partn1,partn2,partn3,cosmv2,eegrg,eeejpp,cosmv3} and references therein. We
obtain a system of nonlinear PDEs which can be represented in effective
Einstein form,
\begin{equation}
\widehat{\mathbf{R}}_{\mu \nu }=\mathbf{\Upsilon }_{\mu \nu },  \label{mfeq}
\end{equation}%
where the right effective source is parameterized
\begin{equation}
\mathbf{\Upsilon }_{\mu \nu }=~^{F}\mathbf{\Upsilon }_{\mu \nu }+~^{m}%
\mathbf{\Upsilon }_{\mu \nu }+~\overline{\mathbf{\Upsilon }}_{\mu \nu }.
\label{sourcc}
\end{equation}%
Let us explain how three terms in this source are defined. The functionals $%
\mathbf{F}(\widehat{\mathbf{R}})$ and \newline
$\ ^{1}\mathbf{F}(\widehat{\mathbf{R}}):=d\mathbf{F}(\widehat{\mathbf{R}})/d%
\widehat{\mathbf{R}}$ determine an energy-momentum tensor,
\begin{equation}
\ \ ^{F}\mathbf{\Upsilon }_{\mu \nu }=(\frac{\mathbf{F}}{2~^{1}\mathbf{F}}-%
\frac{\widehat{\mathbf{D}}^{2}\ ^{1}\mathbf{F}}{~^{1}\mathbf{F}})\mathbf{g}%
_{\mu \nu }+\frac{\widehat{\mathbf{D}}_{\mu }\widehat{\mathbf{D}}_{\nu }\
^{1}\mathbf{F}}{~^{1}\mathbf{F}}.  \label{fsourc}
\end{equation}%
The source for the scalar matter fields can be computed in standard form,
\begin{equation}
~^{m}\mathbf{\Upsilon }_{\mu \nu }=\frac{1}{2M_{P}^{2}}\ ^{m}\mathbf{T}%
_{\alpha \beta },  \label{msourc}
\end{equation}
and the holonomic contributions from LQG, $~\overline{\mathbf{\Upsilon }}%
_{\mu \nu }$ (\ref{lqgcor}), will be defined in subsection \ref{sseclqc}. We
shall be able to find, in explicit form, exact solutions for the system (\ref%
{mfeq}) for any source (\ref{sourcc}), which via frame transforms $\mathbf{%
\Upsilon }_{\mu \nu }=e_{\ \mu }^{\mu ^{\prime }}e_{\ \nu }^{\nu ^{\prime }}%
\mathbf{\Upsilon }_{\mu ^{\prime }\nu ^{\prime }}$ can be parameterized into
N--adapted diagonalized form as
\begin{equation}
\mathbf{\Upsilon }_{\ \nu }^{\mu }=diag[\ ~_{h}\mathbf{\Upsilon (}%
x^{i}),~_{h}\mathbf{\Upsilon (}x^{i}),\ \mathbf{\Upsilon (}x^{i},t),~\mathbf{%
\Upsilon (}x^{i},t)].  \label{sources}
\end{equation}%
In these formulas, the generating source functions $~_{h}\Upsilon (x^{i})$
and $\Upsilon (x^{i},t\mathbf{)}$ have to be prescribed in some forms which
will generate exact solutions compatible with observational/ experimental
data.

\subsection{Small parametric deformations for quasi--FLRW metrics}

\label{sspard}

In N--adapted bases, the models of locally anisotropic and inhomogeneous
anamorphic cosmology are characterized by three essential properties during
the smoothing phase:

\begin{enumerate}
\item masses are polarized with a certain dependence on time and space like
coordinates $m\rightarrow \check{m}(x^{i},t)$ and/or $m\rightarrow \check{m}%
(t)$;

\item necessary type combinations of N-adapted Weyl-invariant signatures
incorporating aspects of contracting and expanding locally anisotropic
backgrounds;

\item using nonlinear symmetries of generic off-diagonal solutions $\mathbf{g%
}_{\alpha \beta }$ (\ref{odans}) and considering nonholonomic deformations
on a small parameter $\varepsilon ,$ we can express, via frame transforms,
the cosmological solutions of (\ref{mfeq}), with prescribed sources $\ [~_{h}%
\mathbf{\Upsilon ,}~\mathbf{\Upsilon ]}$, in such a quasi-FLRW form\footnote{%
this term means that for $\varepsilon \to 0$ and any approximation $\widehat{%
a}^{2}(t)$ a standard FLRW metric is generated}
\begin{equation}
ds^{2}=\widehat{a}^{2}(x^{k},t)\mathbf{e}_{\grave{\imath}}\mathbf{e}^{\grave{%
\imath}}-e_{4}\mathbf{e}^{4},  \label{qflrw}
\end{equation}%
for $N_{j}^{3}=n_{j}(x^{k},t)$ and $N_{j}^{4}=w_{j}(x^{k},t),$ where
\begin{eqnarray}
\mathbf{e}^{\grave{\imath}} &=&(dx^{i},\mathbf{e}%
^{3}=dy^{3}+n_{j}(x^{k},t)dx^{j})\simeq (dx^{i},dy^{3}+\varepsilon \chi
_{j}^{3}(x^{k},t)dx^{j}),  \notag \\
\mathbf{e}^{4} &=&dt+w_{j}(x^{k},t)dx^{j}\simeq dt+\varepsilon \chi
_{i}^{4}(x^{k},t)dx^{j}.  \label{enadcb}
\end{eqnarray}%
The locally anisotropic scale coefficient can be considered as isotropic in
certain limits (for additional assumptions on homogeneity), $\widehat{a}%
^{2}(x^{k},t)\simeq \widehat{a}^{2}(t)$ and computed together with effective
polarization functions $\chi _{j}^{3}$ and $\chi _{i}^{4}$ all encoding data
on possible nonlinear generic off-diagonal interactions, QC and/or LQG
contributions. In next section, we shall prove how such values can be
computed for certain classes of generic off-diagonal exact solutions in MGTs
and GR.
\end{enumerate}

Using the effective scale factor $\widehat{a}^{2}$ from (\ref{qflrw}), we
can introduce the respective effective and locally anisotropically polarized
Hubble parameter,
\begin{equation}
\widehat{H}:=e_{4}(\ln \widehat{a})=\partial _{t}(\ln \widehat{a})=(\ln
\widehat{a})^{\ast }.  \label{hfunct}
\end{equation}%
Considering a new time like coordinate $\check{t},$ for $t=t(x^{i},\check{t}%
) $ and transforming $\sqrt{|h_{4}|}\partial t/\partial \check{t}$ into a
scale factor $\widehat{a}(x^{i},\check{t}),$ we represent (\ref{enadcb}) in
the form%
\begin{eqnarray}
ds^{2} &=&\check{a}^{2}(x^{i},\check{t})[\eta _{i}(x^{k},\check{t}%
)(dx^{i})^{2}+\check{h}_{3}(x^{k},\check{t})(\mathbf{e}^{3})^{2}-(\mathbf{%
\check{e}}^{4})^{2}],  \label{enadcbc} \\
\mbox{ where }\eta _{i} &=&\check{a}^{-2}e^{\psi },\check{a}^{2}\check{h}%
_{3}=h_{3},\mathbf{e}^{3}=dy^{3}+\partial _{k}n~dx^{k},\mathbf{\check{e}}%
^{4}=d\check{t}+\sqrt{|h_{4}|}(\partial _{i}t+w_{i}).  \notag
\end{eqnarray}%
For a small parameter $\varepsilon ,$ with $0\leq \varepsilon <1,$ we the
off--diagonal deformations are given by effective polarization functions
\begin{equation*}
\eta _{i}\simeq 1+\varepsilon \chi _{i}(x^{k},\widehat{t}),\partial
_{k}n\simeq \varepsilon \widehat{n}_{i}(x^{k}),\sqrt{|h_{4}|}\ w_{i}\simeq
\varepsilon \widehat{w}_{i}(x^{k},\widehat{t}).
\end{equation*}%
We can work, for convenience, with both types of nonholonomic $\varepsilon $%
--deformations of FLRW metrics (nonholonomic FLRW models). Such
approximations can be considered after a generic off--diagonal cosmological
solution was constructed in a general form.

\subsection{Effective FLRW geometry for nonholonomic MGTs}

Following a N-adapted variational calculus for MGTs Lagrangians resulting in
respective dynamical equations (see similar holonomic variants in Refs. \cite%
{amoros1,amoros2}), we can construct various models of locally anisotropic
spacetimes \cite{partn1,partn2,partn3,cosmv2,eegrg,eeejpp,cosmv3}.\footnote{%
In our works, we have to elaborate more 'sophisticate' systems of notations
because such geometric modeling of cosmological scenarios and methods of
constructing solutions of PDEs should include various terms with
h-v--splitting; discrete and continuous classical and quantum corrections,
diagonal and off-diagonal terms, different types of connections which were
not considered in other works by other authors. The most important
conventions on our notations are that we use boldface symbols for the spaces
and geometric objects endowed with N--connection structure and that left
labels are abstract ones associated to some classes of geometric/ physical
objects. Right Latin and Greek indices can be abstract ones or transformed
into coordinate indices with possible h- and v-splitting. Unfortunately, it
is not possible to simplify such a system of notations if we follow multiple
purposes related to geometric methods of constructing exact solutions in
gravity and cosmology theories, analysing different phases of anamoprhic
cosmology with generic off-diagonal terms etc.} For $\mathbf{\Upsilon }_{\mu
\nu }=~^{F}\mathbf{\Upsilon }_{\mu \nu }$ in (\ref{mfeq}) and a d--metric (%
\ref{odans}) with diagonal homogeneous approximations, we obtain from (\ref%
{modanamact}) that in the Einstein frame
\begin{eqnarray*}
~^{F}\mathcal{S} &=&M_{P}^{2}\int d^{4}u\sqrt{|\mathbf{g}|}\mathbf{F}(%
\widehat{\mathbf{R}})\rightarrow ~^{EF}\mathcal{S}=M_{P}^{2}\int d^{4}u~^{EF}%
\mathcal{L}, \\
\mbox{ for }~^{EF}\mathcal{L} &=&\overline{a}^{3}\left[ \frac{1}{2}\overline{%
\mathbf{R}}+\frac{1}{2}\left( \frac{\partial \overline{\phi }}{\partial
\overline{t}}\right) ^{2}-V(\overline{\phi })\right] ,\mbox{ where }%
\overline{\mathbf{R}}=6\frac{\partial \overline{H}}{\partial \overline{t}}+12%
\overline{H}^{2}.
\end{eqnarray*}%
In these formulas, $V(\overline{\phi })$ is an effective potential and $%
\overline{a}$ and $\overline{\phi }$ are independent variables defined
correspondingly by
\begin{eqnarray}
\overline{a} &:=&\sqrt{\ ^{1}\mathbf{F}(\widehat{\mathbf{R}})}\widehat{a},\ d%
\overline{t}:=\sqrt{\ ^{1}\mathbf{F}(\widehat{\mathbf{R}})}dt,\frac{\partial
}{\partial \overline{t}}:=\overline{\partial };  \label{efvar} \\
\overline{\phi } &:=&\sqrt{\frac{3}{2}}\ln |\ ^{1}\mathbf{F}(\widehat{%
\mathbf{R}})|,V(\overline{\phi })=\frac{1}{2}\left[ \frac{\widehat{\mathbf{R}%
}}{\ ^{1}\mathbf{F}(\widehat{\mathbf{R}})}-\frac{\mathbf{F}(\widehat{\mathbf{%
R}})}{\left( \ ^{1}\mathbf{F}(\widehat{\mathbf{R}})\right) ^{2}}\right] .
\notag
\end{eqnarray}

Using above variables for the Hamiltonian constraint $\ ^{EF}\mathcal{H}:=%
\overline{\partial }\overline{a}\frac{\partial (~^{EF}\mathcal{L)}}{\partial
\overline{\partial }\overline{a}}+\overline{\partial }\overline{\phi }\frac{%
(\partial ~^{EF}\mathcal{L)}}{\partial \overline{\partial }\overline{\phi }}%
-~^{EF}\mathcal{L}$ and effective density
\begin{equation}
\overline{\rho }:=\frac{1}{2}(\overline{\partial }\overline{\phi })^{2}+V(%
\overline{\phi }),  \label{effd}
\end{equation}%
we express the effective Friedmann equation (in the Einstein frame, it is a
constraint) $3\overline{H}^{2}=\overline{\rho }$ when the dynamics is given
by the conservation law $\overline{\partial }\overline{\rho }=-3\overline{H}(%
\overline{\partial }\overline{\phi })^{2}.$ This dynamics is encoded also in
an effective Raychaudhury equation $2\overline{\partial }\overline{H}=-(%
\overline{\partial }\overline{\phi })^{2}, $ with $(\overline{\partial }%
\overline{\rho })^{2}=3\overline{\rho }(\overline{\partial }\overline{\phi }%
)^{2}.$

\subsection{LQC extensions of MGTs}

\label{sseclqc}LQC corrections to MGTs have been studied in series of works
\cite{zhang1,zhang2,amoros1,amoros2,bamba,amoros}. As standard variables (we
follow our notations (\ref{efvar})), we use $\overline{\beta }:=\overline{%
\gamma }\overline{H},$ where $\overline{\gamma }$ is the Barbero--Immirzi
parameter \cite{barb1,barb2,immirzi}, and the volume $\overline{V}:=%
\overline{a}^{3}.$ For diagonal configurations, the holonomy corrections to
the Friedemann equations are of type
\begin{equation}
\overline{H}^{2}=\frac{\overline{\rho }}{3}(1-\frac{\overline{\rho }}{%
\overline{\rho }_{c}}),  \label{lqgcor}
\end{equation}%
where the critical density $\overline{\rho }_{c}:=2/\sqrt{3}\gamma ^{3}$ is
computed in EF (see \cite{asht0} \ for a status report on different
approaches to LCQ). This formula can be applied for small deformations with
respect to N--adapted frames taking, for simplicity, a function $\overline{%
\rho }(t)$ determining the component $~\overline{\mathbf{\Upsilon }}_{\mu
\nu }$ in (\ref{mfeq}) and (\ref{sourcc}).

In a more general context, we can consider locally anisotropic
configurations with $\overline{\rho }(x^{i},t)$ associated to any $~^{EF}%
\mathcal{H}[\overline{\beta }(x^{i},t),\overline{V}(x^{i},t)],$ with
conjugated Poisson bracket $\{\overline{\beta }(x^{i},t),\overline{V}%
(x^{i},t)\}=\overline{\gamma }/2,$ when $\overline{H}(x^{i},t)=\frac{\sin (%
\sqrt{2\sqrt{3\gamma }}\overline{\beta }(x^{i},t))}{\sqrt{2\sqrt{3\gamma }}}$%
, for a re-scaling in order to have a well--defined quantum theory. We note
that there were formulated different models and inequivalent approaches to
LQG and LQC, see a variant \cite{vlqgdq} which is compatible with
deformation quantization. \ For simplicity, we shall add the term
\begin{equation}
\overline{\mathbf{\Upsilon }}=-\frac{\overline{\rho }^{2}}{3\overline{\rho }%
_{c}}  \label{lqsourc}
\end{equation}
in N-adapted $\overline{\mathbf{\Upsilon }}_{\mu \nu }=diag[\overline{%
\mathbf{\Upsilon }},\overline{\mathbf{\Upsilon }},\overline{\mathbf{\Upsilon
}},\overline{\mathbf{\Upsilon }}],$ see below the formula (\ref{paramsourc}%
), as an additional LQG contribution in the right part of certain
generalized Friedmann equations with a nonlinear re--definition of scalar
field effective density $\frac{\overline{\rho }}{3}\rightarrow \frac{%
\overline{\rho }}{3}(1-\frac{\overline{\rho }}{\overline{\rho }_{c}}).$

\subsection{Nonholonomic Friedmann eqs in anamorphic cosmology with
LQG corrections}

The cosmological models with generic off-diagonal metrics parameterized in
N--adapted form with respect to bases (\ref{enadcb}) are characterized by
two dimensionless quantities (being Weyl-invariant if the homogeneity
conditions are imposed),%
\begin{eqnarray}
~^{m}\Theta := &(\widehat{H}+\overline{H}+\frac{\check{m}^{\ast }}{\check{m}}%
)\check{M}_{P}^{-1}=&\frac{\check{\alpha}_{m}^{\ast }}{\check{\alpha}_{m}%
\check{M}_{P}}\mbox{ for }\check{\alpha}_{m}:=\widehat{a}\check{m}/M_{P}^{0};
\label{dimensionless} \\
~^{Pl}\Theta := &(\widehat{H}+\overline{H}+\frac{\check{M}_{P}^{\ast }}{%
\check{M}_{P}})\check{M}_{P}^{-1}=&\frac{\check{\alpha}_{Pl}^{\ast }}{\check{%
\alpha}_{Pl}\check{M}_{Pl}}\mbox{ for }\check{\alpha}_{Pl}:=\widehat{a}%
\check{M}_{P}/M_{P}^{0}  \notag
\end{eqnarray}%
for $M_{P}^{0}$ being the value of the reduced Planck mass in the frame
where it does not depend on time. These values distinguish respectively such
cosmological models (see details in \cite{sht1a,sht2a} but for holonomic
structures):%
\begin{equation}
\begin{tabular}{lllll}
&  & $\mbox{ anamorphosis }$ & $\mbox{ inflation }$ & $\mbox{ ekpyrosis }$
\\
&  &  &  &  \\
$~^{m}\Theta \mbox{ (background) }$ &  & $<0\mbox{ (contracts)  }$ & $>0%
\mbox{
(expands) }$ & $<0\mbox{ (contracts)  }$ \\
$~^{Pl}\Theta \mbox{ (curvature pert.) }$ &  & $>0\mbox{ (grow) }$ & $>0%
\mbox{
(grow) }$ & $>0\mbox{ (decay) }$%
\end{tabular}
\label{anamcr}
\end{equation}%
Here we note that the priority of the AFDM is that we can consider any
cosmological solution in a MGT or GR and than to write it in N-adapted form
with $\varepsilon $--deformations. This allows us to compute all physical
important values like $~^{m}\Theta $ and $~^{Pl}\Theta $ and analyse if and
when an anamporhic phase is possible. We note that $\ ^{m}\Theta $ is
negative, for instance, as in modified ekpyrotic models, but $~^{Pl}\Theta $
is positive as in locally anisotropic inflationary models. In such theories,
the effective $\check{m}$ and $\check{M}_{P}$ are determined by certain QC
and/or LQC configurations.

Reproducing in N--adapted frames for d--metrics of type (\ref{qflrw}) the
calculus presented in Appendix A (with Einstein and Jordan nonholonomic
frame representations) of \cite{sht1a}, we obtain respectively such a
version of locally anisotropic and inhomogeneous first and second Friedmann
equations,%
\begin{eqnarray}
3(~^{m}\Theta )^{2} &=&\left[ \frac{\ ^{A}\rho +\ ^{m}\rho +\ ^{A}\rho /%
\sqrt{\overline{\rho }_{c}}}{\check{M}_{P}^{4}}-(\frac{\check{m}}{\check{M}%
_{P}})^{2}\frac{\kappa }{\check{\alpha}_{m}^{2}}+(\frac{\check{m}}{\check{M}%
_{P}})^{6}\frac{\sigma ^{2}}{\check{\alpha}_{m}^{6}}\right] \left[
1-\partial (\frac{\check{m}}{\check{M}_{P}})/\partial \ln \check{\alpha}_{m}%
\right] ^{-2},  \notag \\
(~^{Pl}\Theta )^{\ast } &=&-(\ ^{A}\rho +\ ^{m}\rho +\ ^{A}\rho /\sqrt{%
\overline{\rho }_{c}})/2\check{M}_{P}^{3}.  \label{12friedmann}
\end{eqnarray}%
In these formulas, $K(\phi ):=[\frac{3}{2}(f_{,\phi })^{2})+\kappa (\phi
)f(\phi )]/f^{2}(\phi )$\ and the values (energy density, pressure)
\begin{equation*}
2(M_{P}^{0})^{4}\ (\ ^{A}\rho ):=K(\phi )(\phi ^{\ast })^{2}+\check{M}%
_{P}^{4}~^{J}V(\phi )/f^{2}(\phi ),\ 2(M_{P}^{0})^{4}\ (\ ^{A}p):=K(\phi
)(\phi ^{\ast })^{2}-\check{M}_{P}^{4}~^{J}V(\phi )/f^{2}(\phi ),
\end{equation*}%
are determined by coefficients in (\ref{modanamact1}), for $\kappa
=(+1,0,-1) $ being the spacial curvature, and the constant $\sigma ^{2}$
should be considered if we try to limit the background cosmology to that
described by a homogeneous and anisotropic Kasner-like metric (see formula
(A.5) in \cite{sht1a}). For simplicity, we shall consider in this work $%
\sigma ^{2}=1$ even MGTs can contain certain locally anisotropic
configurations.

Finally, we note that we can identify $\ ^{A}\rho $ with $\overline{\rho }$ (%
\ref{effd}) for F--modified gravity theories.

\section{Off-Diagonal Anamorphic Cosmology in MGT and LQG}

\label{sectodancosm} Applying the anholonomic frame deformation method,
AFDM, we can construct various classes of off-diagonal and diagonal
cosmological solutions of (modified) gravitational field equations (\ref%
{mfeq}). After the metric, frame and connection structure, and the effective
sources (\ref{sourcc}), have been parameterized in N-adapted form, we can
select necessary type diagonal or off-diagonal configurations, consider
small parameter decompositions, and approximate the generating/integration
functions to some constant values compatible with observational data. We do
not repeat that geometric formalism and refer readers to Refs. \cite%
{gvvafdm,vcosmsol1,vanp,cosmv2,cosmv3} for details on AFDM and applications
in modern cosmology. The purpose of this section is to state the conditions
for the generating functions and (effective) sources and quantum corrections
which describe quasiperiodic/ aperiodic quasicrystal, QC, like cosmological
structures. There are used necessary type quadratic line elements for
general solutions found in in the mentioned references and the partner
papers \cite{partn1,partn2,partn3}.

\subsection{Generating functions encoding QC like MGT and LQG corrections}

The metrics for off-diagonal locally anisotropic and inhomogeneous
cosmological spacetimes are defined as solutions, with nonholonomically
induced torsion and Killing symmetry on $\partial /\partial y^{3}$.
\footnote{%
For simplicity, in this work we do not consider more general classes of
solutions with generic dependence on all spacetime coordinates and do
analyze the details how Levi-Civita, LC, configurations can be extracted by
solving additional nonholonomic constraints, see \cite%
{gvvafdm,vcosmsol1,vanp,cosmv2,cosmv3} and references therein.} Via
nonholonomic frame transforms, such metrics can be always written in a
coordinate basis, $\mathbf{g}=g_{\alpha \beta }(x^{k},t)du^{\alpha }\otimes
du^{\beta }$, and/or in N--adapted form (\ref{odans}), {\small
\begin{eqnarray}
ds^{2} &=&g_{ij}dx^{i}dx^{j}+\{h_{3}[dy^{3}+(\ _{1}n_{k}+\ _{2}n_{k}\int dt%
\frac{(\partial _{t}\Psi )^{2}}{\ \Upsilon ^{2}|h_{3}|^{5/2}})dx^{k}]^{2}-(%
\frac{\partial _{t}\Psi }{2\Upsilon |h_{3}|^{1/2}})^{2}\ [dt+\frac{\partial
_{i}\Psi }{\ \partial _{t}\Psi }dx^{i}]^{2}\},  \label{tsoluta} \\
h_{3} &=&-\partial _{t}(\Psi ^{2})/\Upsilon ^{2}\left(
h_{3}^{[0]}(x^{k})-\int dt\partial _{t}(\Psi ^{2})/4\Upsilon \right) .
\label{h3}
\end{eqnarray}%
} In this formula, $g_{ij}=\delta _{ij}e^{\ \psi (x^{k})}$ and $\
_{1}n_{k}(x^{i}),\ _{2}n_{k}(x^{i})$ and $h_{a}^{[0]}(x^{k})$ are
integration functions, see details in Appendix \ref{appendixsect}. The
coefficient $h_{3},$ or $\Psi (x^{i},t),$ is the generating function%
\footnote{%
we note that such solutions are defined in explicit form by coefficients of (%
\ref{odans}) computed in this form (see sketch of proofs in Appendix \ref{appendixsect}):
\begin{eqnarray*}
g_{i} &=&e^{\ \psi (x^{k})}\mbox{ is a solution of }\psi ^{\bullet \bullet
}+\psi ^{\prime \prime }=2~\ _{h}\Upsilon ; \\
\ g_{3} &=&h_{3}=-\partial _{t}(\Psi ^{2})/\Upsilon ^{2}\left(
h_{3}^{[0]}(x^{k})-\int dt\partial _{t}(\Psi ^{2})/4\Upsilon \right) ;\
g_{4}=h_{4}^{[0]}(x^{k})-\int dt\partial _{t}(\Psi ^{2})/\ 4\Upsilon ; \\
N_{k}^{3} &=&n_{k}(x^{i},t)=\ _{1}n_{k}(x^{i})+\ _{2}n_{k}(x^{i})\int
dt(\partial _{t}\Psi )^{2}/\Upsilon ^{2}\left\vert h_{3}^{[0]}(x^{i})-\int
dt\ \partial _{t}(\Psi ^{2})/4\Upsilon \right\vert ^{\frac{5}{2}};\
N_{i}^{4}=w_{i}(x^{k},t)=\frac{\partial _{i}\ \Psi }{\partial _{t}\Psi }.
\end{eqnarray*}%
} and the generating $h$- and $v$--sources (see (\ref{sourcc}) and (\ref%
{sources})) are given by terms of effective gravity modifications, matter
field and LQG contributions,
\begin{equation}
\ _{h}\Upsilon (x^{i})=~_{h}^{F}\mathbf{\Upsilon }(x^{i})+~_{h}^{m}\mathbf{%
\Upsilon }(x^{i})+~_{~h}\overline{\mathbf{\Upsilon }}(x^{i})\mbox{ and }%
\mathbf{\Upsilon }(x^{i},t)=~^{F}\mathbf{\Upsilon }(x^{i},t)+~^{m}\mathbf{%
\Upsilon }(x^{i},t)+~\overline{\mathbf{\Upsilon }}(x^{i},t).
\label{paramsourc}
\end{equation}

Off-diagonal metrics of type (\ref{tsoluta}) posses an important nonlinear
symmetry, which allows us to re-define the generating function and
generating source
\begin{eqnarray}
(\Psi ,\ \Upsilon ) &\leftrightarrow &(\Phi ,\Lambda =const),\mbox{ when }%
\Upsilon (x^{k},t)\rightarrow \ \Lambda ,\mbox{ for }  \label{nonlsym} \\
\Phi ^{2} &=&\Lambda \int dt\Upsilon ^{-1}\partial _{t}(\Psi ^{2})%
\mbox{ and
}\Psi ^{2}=\Lambda ^{-1}\int dt\Upsilon \partial _{t}(\Phi ^{2}),  \notag
\end{eqnarray}%
by introducing an effective cosmological constant $\Lambda $ as a source and
the functional $\Phi (\Lambda ,\Psi ,\ \Upsilon )$ as a new generating
function. This property can be proven by considering the relation $\Lambda
\partial _{t}(\Psi ^{2})=\Upsilon \partial _{t}(\Phi ^{2})$ in above
formulas for the d-metric. We can consider that nonlinear generic
off-diagonal interactions on MGTs may induce an effective cosmological
constant with splitting, $\Lambda =\ \ ^{F}\Lambda +\ ^{m}\Lambda +\
\overline{\Lambda }$. The terms of this sum are determined respectively by
modifications of GR resulting in $\ \ ^{F}\Lambda ;$ by nonlinear
interactions of matter (i.e. scalar field $\phi $) resulting in $\
^{m}\Lambda ;$ and by an effective $\ \overline{\Lambda }$ associated to
holonomy modifications from LQG. Technically, it is more convenient to work
with some data $(\Phi ,\Lambda )$ for generating solutions and then to
redefine the formulas in terms of generating function and generalized source
$(\Psi ,\ \Upsilon ).$ We can also extract torsionless cosmological
configurations\footnote{\label{fngr} The nonholonomically induced torsion of
solutions (\ref{tsoluta}) can be constrained to be zero by choosing certain
subclasses of generating functions and sources. We have to consider a
subclass of generating functions and sources when for $\Psi =\check{\Psi}%
(x^{i},t),\partial _{t}(\partial _{i}\check{\Psi})=\partial _{i}(\partial
_{t}\check{\Psi})$ and $\Upsilon (x^{i},t)=\Upsilon \lbrack \check{\Psi}]=%
\check{\Upsilon},$ or $\Upsilon =const.$ Then, we can introduce functions $%
\check{A}(x^{i},t)$ and $n(x^{k})$ subjected to the conditions that $w_{i}=%
\check{w}_{i}=\partial _{i}\check{\Psi}/\partial _{t}\check{\Psi}=\partial
_{i}\check{A}$ and $n_{k}=\check{n}_{k}=\partial _{k}n(x^{i}).$ Such
assumptions are considered in order to simplify the formulas for
cosmological solutions (see details in Refs. \cite%
{gvvafdm,vcosmsol1,vanp,cosmv2,cosmv3}, where the AFDM is applied for
generating more general classes of solutions depending on all spacetime
coordinates. We obtain a quadratic line element defining generic
off-diagonal LC-configurations (a proof is sketched at the end of Appendix \ref{appendixsect}),
\begin{equation*}
ds^{2}=\ g_{ij}dx^{i}dx^{j}+\{h_{3}[dy^{3}+(\partial _{k}n)dx^{k}]^{2}-\frac{%
1}{4h_{3}}\left[ \frac{\partial _{t}\check{\Psi}}{\check{\Upsilon}}\right]
^{2}\ [dt+(\partial _{i}\check{A})dx^{i}]^{2}\}.
\end{equation*}%
}.

The generating functions and/or sources can be chosen in such forms that the
cosmological spacetime solutions encode nontrivial gravitational and/or
matter field \ quasicrystal like, QC, configurations and possible additional
LQG effects. We use an additional 3+1 decomposition with spacelike
coordinates $x^{\grave{\imath}}$ (for $\grave{\imath}=1,2,3$), time like
coordinate $y^{4}=t,$ being adapted to another 2+2 decomposition with a
fibration by 3-d hypersurfaces $\widehat{\Xi }_{t},$ see details in \cite%
{vdoublefibr,mtw}. For such configurations, we can consider a canonical
nonholonomically deformed Laplace operator$\ ^{b}\widehat{\Delta }:=(\ ^{b}%
\widehat{D})^{2}=b^{\grave{\imath}\grave{j}}\widehat{D}_{\grave{\imath}}%
\widehat{D}_{\grave{j}}$ (determined by the 3-d part of d-metric) as a
distortion of $\ ^{b}\Delta :=(\ ^{b}\nabla )^{2}.$ Such a value can be
defined and computed on any $\widehat{\Xi }_{t}$ using a d-metric (\ref%
{odans}) and respective 3-d space like projections/ restrictions of $%
\widehat{\mathbf{D}}.$ We chose a subclass of generating functions $\Psi
=\Pi $ subjected to the condition that it is a solution of an evolution
equation (with conserved dynamics) of type
\begin{equation}
\frac{\partial \Pi }{\partial t}=\ ^{b}\widehat{\Delta }\left[ \frac{\delta
\mathcal{F}}{\delta \Pi }\right] =-\ ^{b}\widehat{\Delta }(\Theta \Pi +Q\Pi
^{2}-\Pi ^{3}).  \label{evoleq}
\end{equation}
Such a nonlinear PDE can be derived for a functional defining an effective
free energy
\begin{equation}
\mathcal{F}[\Pi ]=\int \left[ -\frac{1}{2}\Pi \Theta \Pi -\frac{Q}{3}\Pi
^{3}+\frac{1}{4}\Pi ^{4}\right] \sqrt{b}dx^{1}dx^{2}\delta y^{3},
\label{dener}
\end{equation}%
where $b=\det |b_{\grave{\imath}\grave{j}}|$ is the determinant of the 3-d
spacelike metric, $\delta y^{3}=\mathbf{e}^{3}$ and the operator $\Theta $
and parameter $Q$ are defined in the partner works \cite{partn1,partn2}.
Different choices of $\Theta $ and $Q$ induce different classes of
quasiperiodic, aperiodic and/or QC order of corresponding classes of
gravitational solutions. We note that the functional (\ref{dener}) is of
Lyapunov type considered in quasicrystal physics, see \cite%
{achim,barkan,crystalinks} and references therein, and for applications of
geometric flows in modern cosmology and astrophysics, with generalized
Lyapunov-Perelman functionals \cite{vdoublefibr,vanp,vrajssrf}. In this
paper, we do not enter into details how certain QC structures and their
quasiperiodic/ aperiodic deformations can be reproduced in explicit form but
consider that such configurations can always be modelled by some evolution
equations derived for a respective free energy. The generating / integration
functions and parameters should be chosen in certain forms which are
compatible with experimental data.

Let us explain how the quadratic element (\ref{tsoluta}) defines exact
solutions of MGT field equations (\ref{mfeq}). We prescribe the generating
function and sources with respective associated constants, i.e. certain data
for $\Phi (x^{i},t);\ \ ^{F}\Lambda ,\ ^{m}\Lambda ,\ \overline{\Lambda }$
(defining their sum $\Lambda $); $~_{h}^{F}\mathbf{\Upsilon }%
(x^{i}),~_{h}^{m}\mathbf{\Upsilon }(x^{i}),~_{~h}\overline{\mathbf{\Upsilon }%
}(x^{i})$ and $~^{F}\mathbf{\Upsilon }(x^{i},t)$, $\ ^{m}\mathbf{\Upsilon }%
(x^{i},t)$, $\overline{\mathbf{\Upsilon }}(x^{i},t)$ (defining respective
sums$\ _{h}\Upsilon $ and $\Upsilon ).$ Using formulas (\ref{nonlsym}), we
compute the parametric functional dependence $\Pi =\Psi \lbrack \Phi ;\
^{F}\Lambda ,\ ^{m}\Lambda ,\ \overline{\Lambda };\ ^{F}\mathbf{\Upsilon ,}%
~^{m}\mathbf{\Upsilon },~\overline{\mathbf{\Upsilon }}]$ from%
\begin{equation*}
\Pi ^{2}=(\ \ ^{F}\Lambda +\ ^{m}\Lambda +\ \overline{\Lambda })^{-1}\int
dt(~^{F}\mathbf{\Upsilon }+~^{m}\mathbf{\Upsilon }+~\overline{\mathbf{%
\Upsilon }})\partial _{t}(\Phi ^{2}).
\end{equation*}%
As a result, we can find, in explicit form, the coefficients of d-metric (%
\ref{odans}) parameterized in the form (\ref{enadcbc}), for the class of
generic off-diagonal solutions with Killing symmetry on $\partial _{3}$,
\begin{eqnarray}
g_{i} &=&\check{a}^{2}\eta _{i}=e^{\ \psi (x^{k})}\mbox{ is a solution of }%
\psi ^{\bullet \bullet }+\psi ^{\prime \prime }=2~(~_{h}^{F}\mathbf{\Upsilon
}+~_{h}^{m}\mathbf{\Upsilon }+~_{~h}\overline{\mathbf{\Upsilon }});  \notag
\\
\ g_{3} &=&h_{3}(x^{i},t)=\check{a}^{2}\check{h}_{3}(x^{k},\check{t})=-\frac{%
\partial _{t}(\Pi ^{2})}{(~^{F}\mathbf{\Upsilon }+~^{m}\mathbf{\Upsilon }+~%
\overline{\mathbf{\Upsilon }})^{2}\left( h_{3}^{[0]}(x^{k})-\int dt\frac{%
\partial _{t}(\Pi ^{2})}{4(~^{F}\mathbf{\Upsilon }+~^{m}\mathbf{\Upsilon }+~%
\overline{\mathbf{\Upsilon }})}\right) };  \notag \\
\ g_{4} &=&h_{4}(x^{i},t)=-\check{a}^{2}=h_{4}^{[0]}(x^{k})-\int dt\frac{%
\partial _{t}(\Pi ^{2})}{\ 4(~^{F}\mathbf{\Upsilon }+~^{m}\mathbf{\Upsilon }%
+~\overline{\mathbf{\Upsilon }})};  \label{h4genf} \\
N_{k}^{3} &=&n_{k}(x^{i},t)=\ _{1}n_{k}(x^{i})+\ _{2}n_{k}(x^{i})\int dt%
\frac{(\partial _{t}\Pi )^{2}}{(~^{F}\mathbf{\Upsilon }+~^{m}\mathbf{%
\Upsilon }+~\overline{\mathbf{\Upsilon }})^{2}|h_{3}^{[0]}(x^{i})-\int dt\
\frac{\partial _{t}(\Pi ^{2})}{4(~^{F}\mathbf{\Upsilon }+~^{m}\mathbf{%
\Upsilon }+~\overline{\mathbf{\Upsilon }})}|^{\frac{5}{2}}};\   \notag \\
N_{i}^{4} &=&w_{i}(x^{k},t)=\partial _{i}\ \Pi /\partial _{t}\Pi .  \notag
\end{eqnarray}%
We emphasize that these formulas allow, for instance, to "switch off" the
contributions from LQG if we fix $\overline{\Lambda }=0$ and $\overline{%
\mathbf{\Upsilon }}$ but consider nontrivial values for $\ ^{F}\Lambda +\
^{m}\Lambda $ and $~_{h}^{F}\mathbf{\Upsilon }+~_{h}^{m}\mathbf{\Upsilon .}$

The values $(\ _{h}\Upsilon ,\Upsilon )$ define certain nonholonomic
constraints on the sources and dynamics of (effective) matter fields and
quantum corrections which allows us to integrate a system of nonlinear PDEs
in explicit form and with decoupled $h$- $v$--cosmological evolution in
certain N--adapted systems of reference. In explicit form, we compute using
coefficients of $\widehat{\mathbf{D}}$ for a class of solutions (\ref%
{tsoluta}). At the next step, it is possible to compute $~^{F}\mathbf{%
\Upsilon }_{\mu \nu }$ (\ref{fsourc}),$~^{m}\mathbf{\Upsilon }_{\mu \nu }~$(%
\ref{msourc}) and $\overline{\mathbf{\Upsilon }}_{\mu \nu }$ (\ref{lqsourc})
for arbitrary physically motivated values of F--modifications and solutions
for scalar field $\phi .$ For instance, we generate physically motivated
solutions by considering $\varepsilon $-parametric deformations (\ref{qflrw}%
) of some well defined cosmological solutions in GR or other type MGT, see
examples \cite{vcosmsol1,cosmv2,cosmv3}. For such small off diagonal locally
anisotropic deformations, we have to chose $\widehat{a}^{2}(x^{k},t)$ and $%
\chi _{j}^{a}(x^{k},t)$ to be compatible with experimental gravity and
observation cosmology data.

Other important examples with redefinition and/or prescription of the
generating function and source are those when the integration functions in a
class of metrics (\ref{tsoluta}) are stated to be some constants and, for
instance, $\Phi (x^{i},t)\simeq \Phi (t),$ which results in some data $(\Pi
(t),\ \Upsilon (t))$ following formulas (\ref{nonlsym}). It is also possible
to work with $\varepsilon $-parametric data $(\Phi (\varepsilon
,x^{i},t),\Lambda ),$ and respective $(\Pi (\varepsilon ,x^{i},t),\ \Upsilon
(\varepsilon ,x^{i},t)),$ resulting formulas (\ref{enadcb}) for quasi-FLRW
metrics (\ref{qflrw}). Here, it should be emphasized that even some further
diagonal approximations with $\widehat{a}^{2}(x^{k},t)$ $\simeq \widehat{a}%
^{2}(t)$ will be considered, we shall generate FLRW metrics encoding
partially some data on nonlinear and/or off-diagonal interactions, MGT terms
and LQG corrections. Such solutions can not be found if we introduce
diagonal homogeneous cosmological ansatz which transform, from the very
beginning, the nonlinear systems of PDEs into some ODEs (related to
gravitational and matter field equations in respective theories of gravity
and cosmology).

\subsection{N-adapted Weyl--invariant quantities for anamorphic phases}

For any generic off-diagonal solution (\ref{tsoluta}), we can compute with
respect to N-adapted the $\check{\alpha}$--coefficients and values $%
~^{m}\Theta $ and $~^{Pl}\Theta $ in (\ref{dimensionless}). Re--writing such
solutions in the form (\ref{enadcbc}), with re-defined time like and space
coordinates and scaling factor $\widehat{a}=\check{a}.$ Here, we note that
we can model nonlinear off-diagonal interactions of gravitational and
(effective) matter field interactions in terms of conventional polarization
functions of fundamental physical constants (such values are introduced by
analogy with electromagnetic interactions in certain classical or quantum
media). Let us denote
\begin{equation}
\check{m}=m_{0}\check{\eta}(x^{i},t)\mbox{ and }\check{M}_{Pl}=M_{Pl}^{0}%
\sqrt{f(\phi )}=M_{Pl}^{0}\eta _{Pl}(x^{i},t),  \label{polconst}
\end{equation}
where $\check{\eta}(x^{i},t)$ and $\eta _{Pl}(x^{i},t)$ are respective
polarization of a particle mass $m_{0}$ and Planck constant $M_{Pl}^{0}.$
The $\check{\alpha}$--coefficients in off-diagonal backgrounds are expressed
$\check{\alpha}_{m}:=\widehat{a}\check{\eta}m_{0}/M_{P}^{0}\mbox{ and }%
\check{\alpha}_{Pl}:=\widehat{a}\eta _{Pl}$.

The values for analyzing the conditions for anamorphic phases of (\ref%
{tsoluta}) are computed
\begin{equation}
~^{m}\Theta \lbrack \Pi ]\ M_{Pl}^{0}\eta _{Pl}:=\widehat{H}+\overline{H}+%
\check{\eta}^{\ast }=(\ln |\widehat{a}\check{\eta}|)^{\ast }\mbox{ and }%
~^{Pl}\Theta \lbrack \Pi ]\ M_{Pl}^{0}\eta _{Pl}:=\widehat{H}+\overline{H}%
+\eta _{Pl}^{\ast }=(\ln |\widehat{a}\eta _{Pl}|)^{\ast },  \label{weylinfad}
\end{equation}%
where the Hubble functions, $\widehat{H}$ (\ref{hfunct}) and $\overline{H}$ (%
\ref{lqgcor}) are considered for (\ref{h4genf}) with $h_{4}=-\check{a}^{2}$
and $\overline{\rho }$ (\ref{effd}),
\begin{equation*}
\widehat{H}=(\ln \widehat{a})^{\ast }=\frac{1}{2}\left( \ln \left\vert
h_{4}^{[0]}(x^{k})-\int dt\frac{\partial _{t}(\Pi ^{2})}{\ 4(~^{F}\mathbf{%
\Upsilon }+~^{m}\mathbf{\Upsilon }+~\overline{\mathbf{\Upsilon }})}%
\right\vert \right) ^{\ast }\mbox{ and }\overline{H}=\sqrt{\left\vert \frac{%
\overline{\rho }}{3}(1-\frac{\overline{\rho }}{\overline{\rho }_{c}}%
)\right\vert }.
\end{equation*}

A generating function $\Pi =\Psi \lbrack \Phi ;\ ^{F}\Lambda ,\ ^{m}\Lambda
,\ \overline{\Lambda };^{F}\mathbf{\Upsilon ,}~^{m}\mathbf{\Upsilon },~%
\overline{\mathbf{\Upsilon }}]$ \ may induce anamorphic cosmological phases
following the conditions (\ref{anamcr}) determined by the data for the
integration function $h_{4}^{[0]}(x^{k});$ effective sources $^{F}\mathbf{%
\Upsilon ,}~^{m}\mathbf{\Upsilon ,}~\overline{\mathbf{\Upsilon }}$ and $%
\overline{\rho }$ contained in the sum $\widehat{H}+\overline{H}.$ The
polarizations $\check{\eta}(x^{i},t)$ and $\eta _{Pl}(x^{i},t)$ modify $%
~^{m}\Theta \lbrack \Pi ]\ $\ and $~^{Pl}\Theta \lbrack \Pi ]\ $\ as follow
from $~$(\ref{weylinfad}). Such values can be used for characterizing
locally anisotropic cosmological models, even the analogs of generalized
Friedmann equations (\ref{12friedmann}) for all types of generating
functions.\footnote{%
We can consider a standard interpretation as in \cite{sht1a,sht2a} for small
$\varepsilon $--deformations in Section \ref{epsilonpar}.} We compute
{\small
\begin{equation*}
\begin{tabular}{lllll}
&  & $\mbox{ anamorphosis }$ & $\mbox{ inflation }$ & $\mbox{ ekpyrosis }$
\\
&  &  &  &  \\
$M_{Pl}^{0}~^{m}\Theta \lbrack \Pi ]=(\ln |\sqrt{|h_{4}[\Pi ]|}\check{\eta}%
|)^{\ast }/\eta _{Pl}$ &  & $<0\mbox{ (contracts)  }$ & $>0%
\mbox{
(expands) }$ & $<0\mbox{ (contracts)  }$ \\
$M_{Pl}^{0}~^{Pl}\Theta \lbrack \Pi ]=(\ln |\sqrt{|h_{4}[\Pi ]|}\eta
_{Pl}|)^{\ast }/\eta _{Pl}$ &  & $>0\mbox{ (grow) }$ & $>0\mbox{
(grow) }$ & $>0\mbox{ (decay) }$%
\end{tabular}%
\end{equation*}%
} Such conditions impose additional nonholonomic constraints on generating
functions, sources and integration functions and constants which induce QC
structures as follow from (\ref{dener}).

\subsection{Cosmological QCs for (effective) matter fields and LQG}

Quasiperiodic cosmological structures can be induced by nonholonomic
distributions of (effective) matter fields sources and quantum corrections.

\subsubsection{Effective QC matter fields from MGT}

Let us consider an effective scalar field $\overline{\phi }:=\sqrt{\frac{3}{2%
}}\ln |\ ^{1}\mathbf{F}(\widehat{\mathbf{R}})|$ with nonlinear scalar
potential $V(\overline{\phi })=\frac{1}{2}[\widehat{\mathbf{R}}/\ ^{1}%
\mathbf{F}(\widehat{\mathbf{R}})-\mathbf{F}(\widehat{\mathbf{R}})/\left( \
^{1}\mathbf{F}(\widehat{\mathbf{R}})\right) ^{2}]$ determined by a
modification of GR, see (\ref{efvar}). This results in an effective matter
density $\overline{\rho }:=\frac{1}{2}(\overline{\partial }\overline{\phi }%
)^{2}+V(\overline{\phi })$ and respective $~^{EF}\mathcal{L}.$ Considering
that $V(\overline{\phi })$ is chosen in a form that $\overline{\phi }=\
^{qc}\phi $ (the label $qc$ emphasises modeling a QC structure) is a
solution of
\begin{equation*}
\frac{\partial (\ ^{qc}\phi )}{\partial t}=\ ^{b}\widehat{\Delta }\left[
\frac{\delta (\ ^{qc}\mathcal{F)}}{\delta (\ ^{qc}\phi )}\right] =-\ ^{b}%
\widehat{\Delta }[\Theta \ ^{qc}\phi +Q(\ ^{qc}\phi )^{2}-(\ ^{qc}\phi )^{3}]
\end{equation*}%
with effective free energy $\ ^{qc}\mathcal{F}[\ ^{qc}\phi ]=\int \left[ -%
\frac{1}{2}(\ ^{qc}\phi )\Theta (\ ^{qc}\phi )-\frac{Q}{3}(\ ^{qc}\phi )^{3}+%
\frac{1}{4}(\ ^{qc}\phi )^{4}\right] \sqrt{b}dx^{1}dx^{2}\delta y^{3}$. This
induces an effective matter source of type (\ref{fsourc}),$\ $when$\ \ ^{qc}%
\mathbf{\Upsilon }_{\mu \nu }=\ ^{F}\mathbf{\Upsilon }_{\mu \nu }[\
^{qc}\phi ]=diag(\ _{h}^{qc}\Upsilon ,\ ^{qc}\Upsilon )$ is taken for an
energy momentum tensor $\ \ ^{F}\mathbf{T}_{\mu \nu }$ computed in standard
form for a QC-field $\ ^{qc}\phi .$

We conclude that $F$--modifications of GR can induce QC locally anisotropic
configurations via effective matter field sources if the scalar potential is
determined by a corresponding class of nonlinear interactions and associated
free energy $\ ^{qc}\mathcal{F}.$ Such cosmologies for QC-modified gravity
are described by N--adapted coefficients {\small
\begin{eqnarray}
g_{i} &=&\check{a}^{2}\eta _{i}=e^{\ \psi (x^{k})}\mbox{ is a solution of }%
\psi ^{\bullet \bullet }+\psi ^{\prime \prime }=2~~_{h}^{qc}\mathbf{\Upsilon
};  \label{ex1} \\
\ g_{3} &=&h_{3}(x^{i},t)=\check{a}^{2}\check{h}_{3}(x^{k},\check{t})=-\frac{%
\partial _{t}[\Psi ^{2}(\ ^{qc}\phi )]}{(~^{qc}\mathbf{\Upsilon })^{2}\left(
h_{3}^{[0]}(x^{k})-\int dt\frac{\partial _{t}[\Psi ^{2}(\ ^{qc}\phi )]}{4(~\
^{qc}\Upsilon )}\right) };  \notag \\
\ g_{4} &=&h_{4}(x^{i},t)=-\check{a}^{2}=h_{4}^{[0]}(x^{k})-\int dt\frac{%
\partial _{t}[\Psi ^{2}(\ ^{qc}\phi )]}{\ 4(~\ ^{qc}\Upsilon )};  \notag \\
N_{k}^{3} &=&n_{k}(x^{i},t)=\ _{1}n_{k}(x^{i})+\ _{2}n_{k}(x^{i})\int dt%
\frac{[\partial _{t}\Psi (\ ^{qc}\phi )]^{2}}{(~^{qc}\mathbf{\Upsilon }%
)^{2}|h_{3}^{[0]}(x^{i})-\int dt\ \frac{\partial _{t}[\Psi ^{2}(\ ^{qc}\phi
)]}{4(~^{qc}\mathbf{\Upsilon })}|^{\frac{5}{2}}};\   \notag \\
N_{i}^{4} &=&w_{i}(x^{k},t)=\partial _{i}\ \Psi \lbrack \ ^{qc}\phi
]/\partial _{t}\Psi \lbrack \ ^{qc}\phi ].  \notag
\end{eqnarray}%
} A d-metric (\ref{odans}) with such coefficients describes a cosmological
spacetime encoding "pure" modified gravity contributions. The functional $%
\Psi ^{2}(\ ^{qc}\phi )$ has to be prescribed in a form reproducing
observational data. Considering additional sources for matter fields and
quantum corrections, we can model quasiperiodic and/or aperiodic structures
of different scales and resulting from different sources.

The values necessary for analyzing the conditions for anamorphic phases
induced by QC matter fields from MGT as cosmological spacetimes (\ref{ex1})
are computed
\begin{eqnarray*}
&& \ ^{m}\Theta \lbrack \Psi (\ ^{qc}\phi )]\ M_{Pl}^{0}\eta _{Pl}:=\widehat{%
H}+\check{\eta}^{\ast }=(\ln |\widehat{a}\check{\eta}|)^{\ast }\mbox{ and }%
~^{Pl}\Theta \lbrack \Psi (\ ^{qc}\phi )]\ M_{Pl}^{0}\eta _{Pl}:=\widehat{H}%
+\eta _{Pl}^{\ast }=(\ln |\widehat{a}\eta _{Pl}|)^{\ast } \\
&& \mbox{ where } \widehat{H}=\frac{1}{2}\left( \ln |h_{4}^{[0]}(x^{k})-\int
dt\frac{\partial _{t}[\Psi ^{2}(\ ^{qc}\phi )]}{\ 4(~\ ^{qc}\Upsilon )}%
|\right) ^{\ast }.
\end{eqnarray*}%
A generating function $\Psi \lbrack \ ^{qc}\phi ]$ \ may induce anamorphic
cosmological phases following the conditions {\small
\begin{equation*}
\begin{tabular}{lllll}
&  & $\mbox{ anamorphosis }$ & $\mbox{ inflation }$ & $\mbox{ ekpyrosis }$
\\
&  &  &  &  \\
$M_{Pl}^{0}~^{m}\Theta \lbrack \ ^{qc}\phi ]=(\ln |\sqrt{|h_{4}[\ ^{qc}\phi
]|}\check{\eta}|)^{\ast }/\eta _{Pl}$ &  & $<0\mbox{ (contracts)  }$ & $>0%
\mbox{
(expands) }$ & $<0\mbox{ (contracts)  }$ \\
$M_{Pl}^{0}~^{Pl}\Theta \lbrack \ ^{qc}\phi ]=(\ln |\sqrt{|h_{4}[\ ^{qc}\phi
]|}\eta _{Pl}|)^{\ast }/\eta _{Pl}$ &  & $>0\mbox{ (grow) }$ & $>0%
\mbox{
(grow) }$ & $>0\mbox{ (decay) }$%
\end{tabular}%
\end{equation*}%
} Such conditions impose additional nonholonomic constraints on
modifications of gravity via \newline F--functionals and generating function $\Psi
\lbrack \ ^{qc}\phi ]$ and source $\ ^{qc}\Upsilon $ and integration
functions. We do not consider quantum contributions in generating QCs and
the mass $m_{0}$ is taken by a point particle.

\subsubsection{Nonhomogeneous QC like scalar fields}

For interactions of a scalar field $\ \phi =\ ^{m}\phi $ with mass $m$ and $%
~^{m}\mathbf{\Upsilon }_{\mu \nu }=(2M_{P})^{-2}\ ^{m}\mathbf{T}_{\alpha
\beta }$ (\ref{msourc}) parameterized in N-adapted form, $\ ^{qm}\mathbf{%
\Upsilon }_{\mu \nu }=\ ^{m}\mathbf{\Upsilon }_{\mu \nu }[\ ^{m}\phi
]=diag(\ \ _{h}^{qm}\Upsilon ,\ ^{qm}\Upsilon )$, we can generate QC like
configurations by this class of solutions, {\small
\begin{eqnarray}
g_{i} &=&\check{a}^{2}\eta _{i}=e^{\ \psi (x^{k})}\mbox{ is a solution of }%
\psi ^{\bullet \bullet }+\psi ^{\prime \prime }=2~~_{h}^{qm}\mathbf{\Upsilon
};  \label{ex2} \\
\ g_{3} &=&h_{3}(x^{i},t)=\check{a}^{2}\check{h}_{3}(x^{k},\check{t})=-\frac{%
\partial _{t}[\Psi ^{2}(\ ^{m}\phi )]}{(~^{qm}\mathbf{\Upsilon })^{2}\left(
h_{3}^{[0]}(x^{k})-\int dt\frac{\partial _{t}[\Psi ^{2}(\ ^{m}\phi )]}{4\Psi
(~\ ^{qm}\Upsilon )}\right) };  \notag \\
\ g_{4} &=&h_{4}(x^{i},t)=-\check{a}^{2}=h_{4}^{[0]}(x^{k})-\int dt\frac{%
\partial _{t}[\Psi ^{2}(\ ^{m}\phi )]}{\ 4(~\ ^{qm}\Upsilon )};  \notag \\
N_{k}^{3} &=&n_{k}(x^{i},t)=\ _{1}n_{k}(x^{i})+\ _{2}n_{k}(x^{i})\int dt%
\frac{[\partial _{t}\Psi (\ ^{m}\phi )]^{2}}{(~^{qm}\mathbf{\Upsilon }%
)^{2}|h_{3}^{[0]}(x^{i})-\int dt\ \frac{\partial _{t}[\Psi ^{2}(\ ^{m}\phi )]%
}{4\Psi (~^{qm}\mathbf{\Upsilon })}|^{\frac{5}{2}}};\   \notag \\
N_{i}^{4} &=&w_{i}(x^{k},t)=\partial _{i}\ \Psi \lbrack \ ^{m}\phi
]/\partial _{t}\Psi \lbrack \ ^{m}\phi ].  \notag
\end{eqnarray}%
} We can consider additional constraints for zero torsion configurations
which results in cosmological solutions in GR. Such off-diagonal metrics are
determined by QC like matter distributions if
\begin{equation*}
\frac{\partial (\ ^{m}\phi )}{\partial t}=\ ^{b}\widehat{\Delta }\left[
\frac{\delta (\ ^{qm}\mathcal{F)}}{\delta (\ ^{m}\phi )}\right] =-\ ^{b}%
\widehat{\Delta }[\Theta \ ^{m}\phi +Q(\ ^{m}\phi )^{2}-(\ ^{m}\phi )^{3}]
\end{equation*}%
with effective free energy $\ ^{qm}\mathcal{F}[\ ^{m}\phi ]=\int \left[ -%
\frac{1}{2}(\ ^{m}\phi )\Theta (\ ^{m}\phi )-\frac{Q}{3}(\ ^{m}\phi )^{3}+%
\frac{1}{4}(\ ^{m}\phi )^{4}\right] \sqrt{b}dx^{1}dx^{2}\delta y^{3}$.

It is possible to model double QC configurations with $\overline{\phi }=\
^{qc}\phi +\ ^{m}\phi ,$ for instance, considering $\ ^{m}\phi $ as a small
modification of $\ ^{qc}\phi $ and effective $\mathcal{F\simeq }\ ^{qc}%
\mathcal{F}[\ ^{qc}\phi ]+\ ^{qm}\mathcal{F}[\ ^{m}\phi ].$ In general, we
do not have an additive law of QC free energies for nonlinear MGT and matter
field interactions. The functional $\Psi \lbrack \ ^{m}\phi ]$ is different
from $\Psi \lbrack \ ^{qc}\phi ].$

The values for anamorphic phases induced by QC matter fields from MGT as
cosmological spacetimes (\ref{ex2}) are computed
\begin{eqnarray*}
~^{m}\Theta \lbrack \Psi (\ ^{qc}\phi ),~\ ^{qm}\Upsilon ]\ M_{Pl}^{0}\eta
_{Pl} &:=&\widehat{H}+\check{\eta}^{\ast }=(\ln |\widehat{a}\check{\eta}%
|)^{\ast }\mbox{ and } \\
~^{Pl}\Theta \lbrack \Psi (\ ^{qc}\phi ),~\ ^{qm}\Upsilon ]\ M_{Pl}^{0}\eta
_{Pl} &:=&\widehat{H}+\eta _{Pl}^{\ast }=(\ln |\widehat{a}\eta _{Pl}|)^{\ast
}
\end{eqnarray*}%
where $\widehat{H}=\widehat{H}=\frac{1}{2}\left( h_{4}^{[0]}(x^{k})-\int dt%
\frac{\partial _{t}[\Psi ^{2}(\ ^{m}\phi )]}{\ 4(~\ ^{qm}\Upsilon )}\right)
^{\ast}$. A generating function $\Psi \lbrack \ ^{qc}\phi ]$ \ may induce
anamorphic cosmological phases following the conditions {\small
\begin{equation*}
\begin{tabular}{lllll}
&  & $\mbox{ anamorphosis }$ & $\mbox{ inflation }$ & $\mbox{ ekpyrosis }$
\\
&  &  &  &  \\
$M_{Pl}^{0}~^{m}\Theta \lbrack \Psi (\ ^{qc}\phi ),~\ ^{qm}\Upsilon ]=(\ln |%
\sqrt{|h_{4}|}\check{\eta}|)^{\ast }/\eta _{Pl}$ &  & $<0%
\mbox{ (contracts)
}$ & $>0\mbox{
(expands) }$ & $<0\mbox{ (contracts)  }$ \\
$M_{Pl}^{0}~^{Pl}\Theta \lbrack \Psi (\ ^{qc}\phi ),~\ ^{qm}\Upsilon ]\
=(\ln |\sqrt{|h_{4}|}\eta _{Pl}|)^{\ast }/\eta _{Pl}$ &  & $>0\mbox{ (grow) }
$ & $>0\mbox{
(grow) }$ & $>0\mbox{ (decay) }$%
\end{tabular}%
\end{equation*}%
} These conditions impose additional nonholonomic constraints on generating
function $\Psi \lbrack \ ^{qc}\phi ]$ and source $\ ^{qm}\Upsilon $ and
integration functions. Quantum contributions are not considered and the
scalar field with QC configurations is with polarization of mass $m_{0}.$

\subsubsection{QC configurations induced by LQG corrections}

Quantum corrections may also result in quasiperiodic/ aperiodic QC like
structures, for instance, if LQG sources of type $\check{\Upsilon}=$ $%
\overline{\mathbf{\Upsilon }}=-\overline{\rho }^{2}[\ ^{q}\phi ]/3\overline{%
\rho }_{c}$ (\ref{lqsourc}) are considered for generating cosmological
solutions. This defines an effective scalar field $\overline{\phi }=\
^{q}\phi $ (the label $q$ emphasizes the quantum nature of such a field).
For LQG and GR, such solutions are of type (see footnote \ref{fngr})
\begin{eqnarray}
ds^{2} &=&\ g_{ij}dx^{i}dx^{j}+\{h_{3}[dy^{3}+(\partial _{k}n)dx^{k}]^{2}-%
\frac{9(\overline{\rho }_{c})^{2}}{4h_{3}}\left[ \partial _{t}\check{\Psi}(\
^{q}\phi )\right] ^{2}\ [dt+(\partial _{i}\check{A})dx^{i}]^{2}\},  \notag \\
h_{3} &=&-9(\overline{\rho }_{c})^{2}\partial _{t}(\check{\Psi}^{2})/%
\overline{\rho }^{4}[\ ^{q}\phi ]\left( h_{3}^{[0]}(x^{k})+3\overline{\rho }%
_{c}\int dt\partial _{t}[\check{\Psi}^{2}(\ ^{q}\phi )]/4\overline{\rho }%
^{2}[\ ^{q}\phi ]\right) .  \label{ex3}
\end{eqnarray}%
The QC structure is generated if $\ ^{q}\phi $ is subjected by the
conditions
\begin{equation*}
\frac{\partial (\ ^{q}\phi )}{\partial t}=\ ^{b}\widehat{\Delta }\left[
\frac{\delta (\ ^{q}\mathcal{F)}}{\delta (\ ^{q}\phi )}\right] =-\ ^{b}%
\widehat{\Delta }[\Theta \ ^{q}\phi +Q(\ ^{q}\phi )^{2}-(\ ^{q}\phi )^{3}]
\end{equation*}%
for effective free energy $\ ^{q}\mathcal{F}[\ ^{q}\phi ]=\int \left[ -\frac{%
1}{2}(\ ^{q}\phi )\Theta (\ ^{q}\phi )-\frac{Q}{3}(\ ^{q}\phi )^{3}+\frac{1}{%
4}(\ ^{q}\phi )^{4}\right] \sqrt{b}dx^{1}dx^{2}\delta y^{3}$. This type of
loop QC configurations can be generated from vacuum gravitational fields.

The values for anamorphic phases for QC structures determined by LQG
corrections of matter fields from MGT as cosmological spacetimes (\ref{ex3})
are computed
\begin{eqnarray*}
~^{m}\Theta \lbrack \check{\Psi}(\ ^{q}\phi )]\ M_{Pl}^{0}\eta _{Pl} &:&=%
\widehat{H}+\check{\eta}^{\ast }=(\ln |\widehat{a}\check{\eta}|)^{\ast }%
\mbox{ and } \\
~^{Pl}\Theta \lbrack \check{\Psi}(\ ^{q}\phi )]\ M_{Pl}^{0}\eta _{Pl} &:&=%
\widehat{H}+\eta _{Pl}^{\ast }=(\ln |\widehat{a}\eta _{Pl}|)^{\ast }
\end{eqnarray*}%
where $\widehat{H}=\ln \widehat{a}=\ln |\frac{3\overline{\rho }_{c}}{2h_{3}}%
\partial _{t}\check{\Psi}(\ ^{q}\phi )|$ is computed for%
\begin{equation*}
h_{4}=\overline{\rho }^{4}[\ ^{q}\phi ]\frac{\partial _{t}\check{\Psi}(\
^{q}\phi )}{8\check{\Psi}}\left( h_{3}^{[0]}(x^{k})+\frac{3\overline{\rho }%
_{c}}{4}\int dt\frac{\partial _{t}[\check{\Psi}^{2}(\ ^{q}\phi )]}{\overline{%
\rho }^{2}[\ ^{q}\phi ]}/\right).
\end{equation*}

Anamorphic cosmological phases are determined following the conditions
\begin{equation*}
\begin{tabular}{lllll}
&  & $\mbox{ anamorphosis }$ & $\mbox{ inflation }$ & $\mbox{ ekpyrosis }$
\\
&  &  &  &  \\
$M_{Pl}^{0}~^{m}\Theta \lbrack \check{\Psi}(\ ^{q}\phi )]=(\ln |\sqrt{|h_{4}|%
}\check{\eta}|)^{\ast }/\eta _{Pl}$ &  & $<0\mbox{ (contracts)  }$ & $>0%
\mbox{
(expands) }$ & $<0\mbox{ (contracts)  }$ \\
$M_{Pl}^{0}~^{Pl}\Theta \lbrack \check{\Psi}(\ ^{q}\phi )]\ =(\ln |\sqrt{%
|h_{4}|}\eta _{Pl}|)^{\ast }/\eta _{Pl}$ &  & $>0\mbox{ (grow) }$ & $>0%
\mbox{
(grow) }$ & $>0\mbox{ (decay) }$%
\end{tabular}%
\end{equation*}%
These conditions impose nonholonomic constraints on generating function $%
\check{\Psi}(\ ^{q}\phi )$ for quantum contributions computed in LQG and for
polarization of mass $m_{0}$ of a point particle.

\subsection{Anamorphic off-diagonal cosmology with QC and LQG structures}

Generic off-diagonal solutions (\ref{tsoluta}) encoding parameterized form
QC structures generated by different type sources considered in (\ref{ex1}),
(\ref{ex2}) and (\ref{ex3}) can be written in the form similar to (\ref%
{enadcbc}) with redefined time coordinate and scaling factor $\widehat{a}=%
\check{a}.$ We obtain
\begin{eqnarray}
ds^{2} &=&\widehat{a}^{2}(x^{i},\check{t})[\eta _{i}(x^{k},\check{t}%
)(dx^{i})^{2}+\check{h}_{3}(x^{k},\check{t})(\mathbf{e}^{3})^{2}-(\mathbf{%
\check{e}}^{4})^{2}],  \label{qcflrw} \\
\mbox{ where }\eta _{i} &=&\check{a}^{-2}e^{\psi },\mathbf{e}%
^{3}=dy^{3}+\partial _{k}n(x^{i})~dx^{k},\mathbf{\check{e}}^{4}=d\check{t}+%
\sqrt{|h_{4}|}(\partial _{i}t+w_{i}),  \notag
\end{eqnarray}%
{\small
\begin{eqnarray*}
\mbox{for } \check{h}_{3} &=&-\partial _{t}(\Psi ^{2})/\check{a}^{2}(~^{qc}%
\mathbf{\Upsilon }+~\ ^{qm}\Upsilon -\overline{\rho }^{2}[\ ^{q}\phi ]/3%
\overline{\rho }_{c})^{2} ( h_{3}^{[0]}(x^{k})-\int dt\frac{\partial
_{t}(\Psi ^{2})}{4(~^{qc}\mathbf{\Upsilon }+~\ ^{qm}\Upsilon -\overline{\rho
}^{2}[\ ^{q}\phi ]/3\overline{\rho }_{c})}) \\
h_{4} &=&-\widehat{a}^{2}(x^{i},t)=h_{4}^{[0]}(x^{k})-\int dt\partial
_{t}(\Psi ^{2})/4(~^{qc}\mathbf{\Upsilon }+~\ ^{qm}\Upsilon -\overline{\rho }%
^{2}[\ ^{q}\phi ]/3\overline{\rho }_{c}), \\
w_{i} &=&\partial _{i}\ \Psi /\partial _{t}\Psi ,
\end{eqnarray*}%
} for a functional $\Psi =\Psi \lbrack \ ^{qc}\phi ,\ ^{m}\phi ,\ ^{q}\phi
]. $ For a hierarchy of coupled three QC cosmological structures, we can
subject such a functional of effective sources to conditions of type
\begin{equation*}
\frac{\partial \Psi }{\partial t}=\ ^{b}\widehat{\Delta }\left[ \frac{\delta
\mathcal{F}}{\delta \Psi }\right] =-\ ^{b}\widehat{\Delta }(\Theta \Psi
+Q\Psi ^{2}-\Psi ^{3}),
\end{equation*}
with a functional for effective free energy $\mathcal{F}[\Psi ]=\int \left[ -%
\frac{1}{2}\Psi \Theta \Psi -\frac{Q}{3}\Psi ^{3}+\frac{1}{4}\Psi ^{4}\right]
\sqrt{b}dx^{1}dx^{2}\delta y^{3}$, written in conventional
integro-functional forms.

The values characterizing anamorphic phases in QC cosmological spacetimes
are computed
\begin{equation*}
~^{m}\Theta \ M_{Pl}^{0}\eta _{Pl}:=\widehat{H}+\overline{H}+\check{\eta}%
^{\ast }=(\ln |\widehat{a}\check{\eta}|)^{\ast }\mbox{ and }~^{Pl}\Theta \
M_{Pl}^{0}\eta _{Pl}:=\widehat{H}+\overline{H}+\eta _{Pl}^{\ast }=(\ln |%
\widehat{a}\eta _{Pl}|)^{\ast }
\end{equation*}%
where the polarized Hubble functions, $\widehat{H}$ (\ref{hfunct}) and $%
\overline{H}$ (\ref{lqgcor}), are taken for the quadratic element (\ref%
{qcflrw})
\begin{equation*}
\widehat{H}=(\ln \widehat{a})^{\ast }=\frac{1}{2}\left( \ln \left\vert
h_{4}^{[0]}(x^{k})-\int dt\frac{\partial _{t}(\Psi ^{2})}{4(~^{qc}\mathbf{%
\Upsilon }+~\ ^{qm}\Upsilon -\overline{\rho }^{2}[\ ^{q}\phi ]/3\overline{%
\rho }_{c})}\right\vert \right) ^{\ast }\mbox{ and }\overline{H}=\sqrt{%
\left\vert \frac{\overline{\rho }}{3}(1-\frac{\overline{\rho }}{\overline{%
\rho }_{c}})\right\vert }.
\end{equation*}

A generating function $\Psi =\Psi \lbrack \Phi ;\ ^{F}\Lambda ,\ ^{m}\Lambda
,\ \overline{\Lambda };^{F}\mathbf{\Upsilon ,}~^{m}\mathbf{\Upsilon },~%
\overline{\mathbf{\Upsilon }}]$ \ may induce anamorphic cosmological phases
following the conditions (\ref{anamcr}). In the case of mixed 3 type QC
structures, the Weyl type anamorphic characteristics are determined also by
the data for the integration function $h_{4}^{[0]}(x^{k});$ effective
sources $^{F}\mathbf{\Upsilon ,}~^{m}\mathbf{\Upsilon ,}~\overline{\mathbf{%
\Upsilon }}$ and $\overline{\rho }$ contained in the sum $\widehat{H}+%
\overline{H}.$ We compute {\small
\begin{equation*}
\begin{tabular}{lllll}
&  & $\mbox{ anamorphosis }$ & $\mbox{ inflation }$ & $\mbox{ ekpyrosis }$
\\
&  &  &  &  \\
$M_{Pl}^{0}~^{m}\Theta \lbrack \Psi ,~^{qc}\mathbf{\Upsilon ,}\
^{qm}\Upsilon ,\overline{\rho }^{2}]=(\ln |\sqrt{|h_{4}|}\check{\eta}%
|)^{\ast }/\eta _{Pl}$ &  & $<0\mbox{ (contracts)  }$ & $>0%
\mbox{
(expands) }$ & $<0\mbox{ (contracts)  }$ \\
$M_{Pl}^{0}~^{Pl}\Theta \lbrack \Psi ,~^{qc}\mathbf{\Upsilon ,}\
^{qm}\Upsilon ,\overline{\rho }^{2}]=(\ln |\sqrt{|h_{4}|}\eta _{Pl}|)^{\ast
}/\eta _{Pl}$ &  & $>0\mbox{ (grow) }$ & $>0\mbox{
(grow) }$ & $>0\mbox{ (decay) }$%
\end{tabular}%
\end{equation*}%
} Such conditions impose additional nonholonomic constraints on generating
functions and all types of sources and integration functions and constants
which induce QC structures.

\section{Small Parametric Anamorphic Cosmological QC and LQG Structures}

\label{epsilonpar}The main goal of this paper is to prove that quasiperiodic
and/or aperiodic (for instance, QC like) structures in MGT with LQG helicity
contributions can be incorporated in a compatible way in the framework of
the anamorphic cosmology \cite{sht1a,sht2a,sht3a,sht4a,sht5a,ijj}. For the
classes of cosmological solutions constructed in general form in previous
section, we can consider a procedure of small $\varepsilon $--deformations
of d-metrics of type (\ref{tsoluta}) with respective N--adapted frames and
connections, see details in \cite%
{vcosmsol1,vdoublefibr,gvvafdm,vanp,vrajssrf,cosmv2,cosmv3} and subsection %
\ref{sspard}. In this section, we show how using $\varepsilon $%
--deformations an off--diagonal \ "prime" metric, $\mathbf{\mathring{g}}%
(x^{i},y^{3},t)$\textbf{\ }(for applications in modern cosmolgoy, this
metric can be diagonalizable under coordinate transforms\footnote{%
we note that in general, $\mathbf{\mathring{g}}$ (\ref{pm}) may not be a
solution of gravitational field equations but it will be nonholonomically
deformed into such solutions.}\textbf{)} into a \ "target" metric, $\mathbf{g%
}(x^{i},y^{3},t).$

\subsection{N-adapted $\protect\varepsilon $--deformations}

We suppose that a "prime " pseudo--Riemannian cosmological metric $\mathbf{%
\mathring{g}}=[\mathring{g}_{i},\mathring{h}_{a},\mathring{N}_{b}^{j}]$ can
be parameterized in the form
\begin{eqnarray}
ds^{2} &=&\mathring{g}_{i}(x^{k},t)(dx^{i})^{2}+\mathring{h}_{a}(x^{k},t)(%
\mathbf{\mathring{e}}^{a})^{2},  \label{pm} \\
\mathbf{\mathring{e}}^{3} &=&dy^{3}+\mathring{n}_{i}(x^{k},t)dx^{i},\mathbf{%
\mathring{e}}^{4}=dt+\mathring{w}_{i}(x^{k},t)dx^{i}.  \notag
\end{eqnarray}%
For instance, some data $(\mathring{g}_{i},\mathring{h}_{a})$ may define a
cosmological solution in MGT or in GR like a FLRW, metric. The target metric
$\mathbf{g=}\ ^{\varepsilon }\mathbf{g}$ for an off-diagonal deformation of
the metric structure, for a small parameter $0\leq \varepsilon \ll 1,$ is
parameterized by N-adapted quadratic elements
\begin{eqnarray}
ds^{2} &=&\overline{\eta }_{i}(x^{k},t)\mathring{g}_{i}(x^{k},t)(dx^{i})^{2}+%
\overline{\eta }_{a}(x^{k},t)\mathring{g}_{a}(x^{k},t)(\mathbf{e}^{a})^{2}
\label{targm} \\
&=&\check{a}^{2}(x^{i},t)[\eta _{i}(x^{k},t)(dx^{i})^{2}+\check{h}%
_{3}(x^{k},t)(\mathbf{e}^{3})^{2}-(\mathbf{\check{e}}^{4})^{2}],  \notag \\
\mathbf{e}^{3} &=&dy^{3}+\ ^{n}\eta _{i}(x^{k},t)\mathring{n}%
_{i}(x^{k},t)dx^{i}=dy^{3}+\partial _{k}n~dx^{k},  \notag \\
\mathbf{e}^{4} &=&dt+\ ^{w}\eta _{i}(x^{k},t)\mathring{w}_{i}(x^{k},t)dx^{i}=%
\mathbf{\check{e}}^{4}=d\check{t}+\sqrt{|h_{4}|}(\partial _{i}t+w_{i}),
\notag
\end{eqnarray}%
with possible re-definitions of coordinates $\check{t}=\check{t}(x^{k},t)$
for $\check{a}^{2}(x^{i},t)\rightarrow \widehat{a}^{2}(x^{i},t)$ and where,
for instance, $\ ^{n}\eta _{i}\mathring{n}_{i}dx^{i}=\ ^{n}\eta _{1}%
\mathring{n}_{1}dx^{1}+\ ^{n}\eta _{2}\mathring{n}_{2}dx^{2}.$ The
polarization functions are $\varepsilon $-deformed following rules adapted
to (\ref{enadcbc}) \ and (\ref{qcflrw}), when
\begin{eqnarray}
\overline{\eta }_{i} &=&\check{\eta}_{i}(x^{k},t)[1+\varepsilon \overline{%
\chi }_{i}(x^{k},t)],\overline{\eta }_{a}=1+\varepsilon \overline{\chi }%
_{a}(x^{k},t)\mbox{ and }  \label{smpolariz} \\
\ ^{n}\eta _{i} &=&1+\varepsilon \ \ ^{n}\chi _{i}(x^{k},t),\ ^{w}\eta
_{i}=1+\varepsilon \ ^{w}\chi _{i}(x^{k},t),  \notag \\
\eta _{i} &\simeq &1+\varepsilon \chi _{i}(x^{k},\widehat{t}),\partial
_{k}n\simeq \varepsilon \widehat{n}_{i}(x^{k}),\sqrt{|h_{4}|}\ w_{i}\simeq
\varepsilon \widehat{w}_{i}(x^{k},\widehat{t}).  \notag
\end{eqnarray}%
Such "double" N--adapted deformations are convenient for generating new
classes of solutions and further physical interpretation of such solutions
with limits of quasi-FLRW metrics to some homogenous diagonal cosmological
metrics.

The target generic off--diagonal cosmological metrics
\begin{equation*}
\mathbf{g=}\ ^{\varepsilon }\mathbf{g=(}\ ^{\varepsilon }g_{i},\
^{\varepsilon }h_{a},\ ^{\varepsilon }N_{b}^{j})=(g_{\alpha }=\eta _{\alpha }%
\mathring{g}_{\alpha },\ ^{n}\eta _{i}n_{i},\ ^{w}\eta _{i}\mathring{w}_{i})%
\mbox{ (\ref{targm}) }\rightarrow \mathbf{\mathring{g}}\mbox{ (\ref{pm}) }%
\mbox{ for }\varepsilon \rightarrow 0,
\end{equation*}
define, for instance, cosmological QC configurations with parametric $%
\varepsilon $-dependence determined by a class of solutions (\ref{tsoluta})
(or any variant of solutions (\ref{h4genf}), (\ref{ex1}), (\ref{ex2}), (\ref%
{ex3}) and (\ref{qcflrw})). The effective $\varepsilon $-polarizations of
constants (see (\ref{polconst})) are written
\begin{equation*}
\check{m}=m_{0}\check{\eta}(x^{i},t)\simeq m_{0}(1+\varepsilon \chi
(x^{i},t))\mbox{ and }\check{M}_{Pl}=M_{Pl}^{0}\sqrt{f(\phi )}%
=M_{Pl}^{0}\eta _{Pl}(x^{i},t)=M_{Pl}^{0}(1+\varepsilon \chi _{Pl}(x^{i},t)),
\end{equation*}%
see formulas (\ref{modanamact}), (\ref{modanamact1}) and (\ref{dimensionless}%
), where $\ ^{A}\rho =$ $\overline{\rho }$ (\ref{effd}) in locally
anisotropic and inhomogeneous first and second Friedmann equations, (\ref%
{12friedmann}).

\subsection{$\protect\varepsilon $--deformations to off-diagonal
cosmological metrics}

The deformations of $h$-components of the cosmological d--metrics are
\begin{equation*}
\ ^{\varepsilon }g_{i}(x^{k})=\mathring{g}_{i}(x^{k},t)\check{\eta}%
_{i}(x^{k},t)[1+\varepsilon \chi _{i}(x^{k},t)]=e^{\psi (x^{k})}
\end{equation*}%
defined by a solution of the 2-d Poisson equation. Considering $\psi =\
^{0}\psi (x^{k})+\varepsilon \ \ ^{1}\psi (x^{k})$ and $\ ~$%
\begin{equation*}
_{h}\mathbf{\Upsilon }(x^{k})=\ _{h}^{0}\Upsilon (x^{k})+\varepsilon \
_{h}^{1}\Upsilon (x^{k})=~_{h}^{F}\mathbf{\Upsilon }(x^{i})+~_{h}^{m}\mathbf{%
\Upsilon }(x^{i})+~_{~h}\overline{\mathbf{\Upsilon }}(x^{i})
\end{equation*}
(in particular, we can take $\ _{h}^{1}\Upsilon ),$ see (\ref{sources}), we
compute the deformation polarization functions%
\begin{equation}
\overline{\chi }_{i}=e^{\ \ _{\backepsilon }^{0}\psi }\ \ ^{1}\psi /%
\mathring{g}_{i}\check{\eta}_{i}\ \ \ _{h}^{1}\Upsilon .  \label{aux2a}
\end{equation}

Let us compute $\varepsilon $--deformations of $v$--components using
formulas for a source $\mathbf{\Upsilon }=~^{F}\mathbf{\Upsilon }+~^{m}%
\mathbf{\Upsilon }+~\overline{\mathbf{\Upsilon }}.\ $We consider
\begin{equation}
\ ^{\varepsilon }h_{3}=h_{3}^{[0]}(x^{k})-\frac{1}{4}\int dt\frac{(\Psi
^{2})^{\ast }}{\ \ \Upsilon }=(1+\varepsilon \overline{\chi }_{3})\mathring{g%
}_{3}\ ;\ ^{\varepsilon }h_{4}=-\frac{1}{4}\frac{(\ \Psi ^{\ast })^{2}}{%
(\Upsilon )^{2}}\left( h_{4}^{[0]}-\frac{1}{4}\int dt\frac{(\Psi ^{2})^{\ast
}}{\Upsilon }\right) ^{-1}=(1+\varepsilon \ \overline{\chi }_{4})\mathring{g}%
_{4},  \label{h34b}
\end{equation}%
when the generation function can also be $\varepsilon $--deformed,
\begin{equation}
\Psi =\ ^{\varepsilon }\Psi =\mathring{\Psi}(x^{k},t)[1+\varepsilon
\overline{\chi }(x^{k},t)].  \label{aux5}
\end{equation}%
Introducing $\ ^{\varepsilon }\Psi $ in (\ref{h34b}), we compute
\begin{equation}
\overline{\chi }_{3}=-\frac{1}{4\mathring{g}_{3}}\int dt\frac{(\mathring{\Psi%
}^{2}\overline{\chi })^{\ast }}{\ \Upsilon }\mbox{ and }\int dt\frac{(%
\mathring{\Psi}^{2})^{\ast }}{\ \Upsilon }=4(h_{3}^{[0]}-\mathring{g}_{3}).
\label{cond2a}
\end{equation}%
We conclude that $\overline{\chi }_{3}$ can be computed for any deformation $%
\overline{\chi }$ in (\ref{aux5}) adapted to a time like oriented family of
2-hypersurfaces $t=t(x^{k}).$ This family given in non-explicit form by $%
\mathring{\Psi}=\mathring{\Psi}(x^{k},t)$ when the integration function $%
h_{3}^{[0]}(x^{k}),$ $\mathring{g}_{3}(x^{k})$ and $(\mathring{\Psi}%
^{2})^{\ast }/\ \Upsilon $ satisfy the conditions (\ref{cond2a}).

Using (\ref{aux5}) and (\ref{h34b}), we get%
\begin{equation*}
\overline{\chi }_{4}=2(\overline{\chi }+\frac{\mathring{\Psi}}{\mathring{\Psi%
}^{\ast }}\overline{\chi }^{\ast })-\overline{\chi }_{3}=2(\overline{\chi }+%
\frac{\mathring{\Psi}}{\mathring{\Psi}^{\ast }}\overline{\chi }^{\ast })+%
\frac{1}{4\mathring{g}_{3}}\int dt\frac{(\mathring{\Psi}^{2}\overline{\chi }%
)^{\ast }}{\Upsilon }.
\end{equation*}%
As a result, we can compute $\overline{\chi }_{3}$ for any data $\left(
\mathring{\Psi},\mathring{g}_{3},\overline{\chi }\right) $ and a compatible
source $\ \Upsilon =\pm \mathring{\Psi}^{\ast }/2\sqrt{|\mathring{g}%
_{4}h_{3}^{[0]}|}.$ Such conditions and (\ref{cond2a}) define a time
oriented family of 2-d hypersurfaces, parameterized by $t=t(x^{k})$ defined
in non-explicit form from
\begin{equation}
\int dt\mathring{\Psi}=\pm (h_{3}^{[0]}-\mathring{g}_{3})/\sqrt{|\mathring{g}%
_{4}h_{3}^{[0]}|}.  \label{cond2b}
\end{equation}

The final step consists of $\varepsilon $--deformations N--connection
coefficients $\ w_{i}=\partial _{i}\Psi /\Psi ^{\ast }$ for nontrivial $\
\mathring{w}_{i}=\partial _{i}\ \mathring{\Psi}/\ \mathring{\Psi}^{\ast },$
which are computed following formulas (\ref{aux5}) and (\ref{smpolariz}), $\
^{w}\chi _{i}=\frac{\partial _{i}(\overline{\chi }\ \mathring{\Psi})}{%
\partial _{i}\ \mathring{\Psi}}-\frac{(\overline{\chi }\ \mathring{\Psi}%
)^{\diamond }}{\mathring{\Psi}^{\diamond }}$. We omit similar computations
of $\varepsilon $--deformations of $n$--coefficients (we omit such details
which are not important if we restrict our research only to
LC-configurations).

Summarizing (\ref{aux2a})-(\ref{cond2b}), we obtain the following formulas
for $\varepsilon $--deformations of a prime cosmological metric (\ref{pm})
into a target cosmological metric: {\small
\begin{eqnarray}
\ \ ^{\varepsilon }g_{i} &=&[1+\varepsilon \overline{\chi }_{i}(x^{k},t)]%
\mathring{g}_{i}\check{\eta}_{i}=[1+\varepsilon e^{\ \ ^{0}\psi }\ \
^{1}\psi /\mathring{g}_{i}\check{\eta}_{i}\ \ \ _{h}^{0}\Upsilon ]\mathring{g%
}_{i}\mbox{ solution of 2-d Poisson eqs )};  \notag \\
\ \ ^{\varepsilon }h_{3} &=&[1+\varepsilon \ \overline{\chi }_{3}]\mathring{g%
}_{3}=\left[ 1-\varepsilon \frac{1}{4\mathring{g}_{3}}\int dt\frac{(%
\mathring{\Psi}^{2}\overline{\chi })^{\ast }}{\ \ \Upsilon }\right]
\mathring{g}_{3};  \notag \\
\ ^{\varepsilon }h_{4} &=&[1+\varepsilon \ \overline{\chi }_{4}]\mathring{g}%
_{4}=\left[ 1+\varepsilon \ \left( 2(\overline{\chi }+\frac{\mathring{\Psi}}{%
\mathring{\Psi}^{\ast }}\overline{\chi }^{\ast })+\frac{1}{4\mathring{g}_{3}}%
\int dt\frac{(\mathring{\Psi}^{2}\overline{\chi })^{\ast }}{\ \Upsilon }%
\right) \right] \mathring{g}_{4};  \label{ersdef} \\
\ ^{\varepsilon }n_{i} &=&[1+\varepsilon \ ^{n}\chi _{i}]\mathring{n}_{i}=%
\left[ 1+\varepsilon \ \widetilde{n}_{i}\int dt\ \frac{1}{\Upsilon ^{2}}%
\left( \overline{\chi }+\frac{\mathring{\Psi}}{\mathring{\Psi}^{\ast }}%
\overline{\chi }^{\ast }+\frac{5}{8}\frac{1}{\mathring{g}_{3}}\frac{(%
\mathring{\Psi}^{2}\overline{\chi })^{\ast }}{\ \Upsilon }\right) \right]
\mathring{n}_{i};  \notag \\
\ \ ^{\varepsilon }w_{i} &=&[1+\varepsilon \ ^{w}\chi _{i}]\mathring{w}_{i}=%
\left[ 1+\varepsilon (\frac{\partial _{i}(\overline{\chi }\ \mathring{\Psi})%
}{\partial _{i}\ \mathring{\Psi}}-\frac{(\overline{\chi }\ \mathring{\Psi}%
)^{\ast }}{\mathring{\Psi}^{\ast }})\right] \mathring{w}_{i}.  \notag
\end{eqnarray}%
} The factor $\ \widetilde{n}_{i}(x^{k})$ is a redefined integration
function.

The quadratic element for such inhomogeneous and locally anisotropic
cosmological spaces with coefficients (\ref{ersdef}) can be written in
N--adapted form
\begin{eqnarray}
ds^{2} &=&\ \ ^{\varepsilon }g_{\alpha \beta }(x^{k},t)du^{\alpha }du^{\beta
}=\ ^{\varepsilon }g_{i}\left( x^{k}\right) [(dx^{1})^{2}+(dx^{2})^{2}]+
\label{riccisoltdef} \\
&&\ \ ^{\varepsilon }h_{3}(x^{k},t)\ [dy^{3}+\ \ \ ^{\varepsilon
}n_{i}dx^{i}]^{2}+\ ^{\varepsilon }h_{4}(x^{k},t)[dt+\ \ ^{\varepsilon
}w_{k}\ (x^{k},t)dx^{k}]^{2}.  \notag
\end{eqnarray}%
Further assumptions on generating and integration functions and source can
be considered in order to find solutions of type $\ ^{\varepsilon }g_{\alpha
\beta }(x^{k},t)\simeq \ ^{\varepsilon }g_{\alpha \beta }(t).$

\subsection{Cosmological $\varepsilon $--deformations with
anamorphic QCs and LQG}

We apply the procedure of $\varepsilon $--deformations described in the
previous subsection in order to generate solutions of type (\ref{qcflrw}).
We prescribe $\overline{\chi }(x^{k},t)$ and $\ \ \ _{h}^{0}\Upsilon (x^{k})$
for any compatible $\left( \mathring{\Psi},\mathring{g}_{3}\right)$ and
source
\begin{equation*}
\mathring{\Psi}^{\ast }=\pm 2\sqrt{|\mathring{g}_{4}h_{3}^{[0]}|}\ (~^{qc}%
\mathbf{\Upsilon }+~\ ^{qm}\Upsilon -\overline{\rho }^{2}[\ ^{q}\phi ]/3%
\overline{\rho }_{c}).
\end{equation*}%
The generated d--metric with coefficients (\ref{ersdef}) is of type (\ref%
{riccisoltdef}) for $\Upsilon =~^{qc}\mathbf{\Upsilon }+~\ ^{qm}\Upsilon -%
\overline{\rho }^{2}[\ ^{q}\phi ]/3\overline{\rho }_{c},$
\begin{eqnarray*}
ds^{2} &=&\ [1+\varepsilon e^{\ \ ^{0}\psi }\ \ ^{1}\psi /\mathring{g}_{i}%
\check{\eta}_{i}\ \ \ _{h}^{0}\Upsilon ]\mathring{g}%
_{i}[(dx^{1})^{2}+(dx^{2})^{2}]+ \\
&&\ \ [1-\varepsilon \frac{1}{4\mathring{g}_{3}}\int dt\frac{(\mathring{\Psi}%
^{2}\overline{\chi })^{\ast }}{\ \ \Upsilon }]\mathring{g}_{3}\ \left[
dy^{3}+\ [1+\varepsilon \ \widetilde{n}_{i}\int dt\ \frac{1}{\Upsilon ^{2}}%
\left( \overline{\chi }+\frac{\mathring{\Psi}}{\mathring{\Psi}^{\ast }}%
\overline{\chi }^{\ast }+\frac{5}{8}\frac{1}{\mathring{g}_{3}}\frac{(%
\mathring{\Psi}^{2}\overline{\chi })^{\ast }}{\ \Upsilon }\right) ]\mathring{%
n}_{i}dx^{i}\right] ^{2}+ \\
&&\lbrack 1+\varepsilon \ (2(\overline{\chi }+\frac{\mathring{\Psi}}{%
\mathring{\Psi}^{\ast }}\overline{\chi }^{\ast })+\frac{1}{4\mathring{g}_{3}}%
\int dt\frac{(\mathring{\Psi}^{2}\overline{\chi })^{\ast }}{\ \Upsilon })]%
\mathring{g}_{4}\left[ dt+\ [1+\varepsilon (\frac{\partial _{i}(\overline{%
\chi }\ \mathring{\Psi})}{\partial _{i}\ \mathring{\Psi}}-\frac{(\overline{%
\chi }\ \mathring{\Psi})^{\ast }}{\mathring{\Psi}^{\ast }})]\mathring{w}%
_{k}dx^{k}\right] ^{2}.
\end{eqnarray*}

Hierarchies of coupled three QC cosmological structures are generated by a
functional \newline $\overline{\chi }=\overline{\chi }[\ ^{qc}\phi ,\ ^{m}\phi ,\
^{q}\phi ]$ subjected to conditions of type
\begin{equation*}
\frac{\partial \overline{\chi }}{\partial t}=\ ^{b}\widehat{\Delta }\left[
\frac{\delta \mathcal{F}}{\delta \Psi }\right] =-\ ^{b}\widehat{\Delta }%
(\Theta \overline{\chi }+Q\overline{\chi }^{2}-\overline{\chi }^{3}),
\end{equation*}%
with functionals for effective free energy $\mathcal{F}[\overline{\chi }%
]=\int \left[ -\frac{1}{2}\overline{\chi }\Theta \overline{\chi }-\frac{Q}{3}%
\overline{\chi }^{3}+\frac{1}{4}\overline{\chi }^{4}\right] \sqrt{b}%
dx^{1}dx^{2}\delta y^{3}$, written in conventional integro-functional forms.
The value
\begin{equation}
\ ^{\varepsilon }h_{4}=-\ ^{\varepsilon }\widehat{a}^{2}(x^{i},t)=[1+%
\varepsilon \ (2(\overline{\chi }+\frac{\mathring{\Psi}}{\mathring{\Psi}%
^{\ast }}\overline{\chi }^{\ast })+\frac{1}{4\mathring{g}_{3}}\int dt\frac{(%
\mathring{\Psi}^{2}\overline{\chi })^{\ast }}{\ \Upsilon })]\mathring{g}_{4},
\label{h4e}
\end{equation}%
with $\mathring{g}_{4}=a(t),$ allows us to compute the Weyl type invariants
characterizing anamporphic phases in QC cosmological spacetimes,
\begin{eqnarray*}
~_{\varepsilon }^{m}\Theta \ M_{Pl}^{0}(1+\varepsilon \chi _{PL}) &:&=%
\widehat{H}+\overline{H}+(1+\varepsilon \chi )^{\ast }=(\ln |\ ^{\varepsilon
}\widehat{a}(1+\varepsilon \chi )|)^{\ast }\mbox{ and }~ \\
_{\varepsilon }^{Pl}\Theta \ M_{Pl}^{0}(1+\varepsilon \chi _{PL}) &:&=%
\widehat{H}+\overline{H}+(1+\varepsilon \chi )^{\ast }=(\ln |\ ^{\varepsilon
}\widehat{a}(1+\varepsilon \chi _{PL})|)^{\ast },
\end{eqnarray*}%
where the $\varepsilon $-polarized Hubble functions, $\ ^{\varepsilon }%
\widehat{H}$ (\ref{hfunct}) and $\ ^{\varepsilon }\overline{H}$ (\ref{lqgcor}%
) are respectively computed for $\ ^{\varepsilon }h_{4}$ {\small
\begin{equation*}
\ ^{\varepsilon }\widehat{H} =(\ln \ ^{\varepsilon }\widehat{a})^{\ast }=%
\frac{1}{2}\left( \ln \left\vert [1+\varepsilon \ (2(\overline{\chi }+\frac{%
\mathring{\Psi}}{\mathring{\Psi}^{\ast }}\overline{\chi }^{\ast })+\frac{1}{4%
\mathring{g}_{3}}\int dt\frac{(\mathring{\Psi}^{2}\overline{\chi })^{\ast }}{%
\ \Upsilon })]\mathring{g}_{4}\right\vert \right) ^{\ast } \mbox{ and }\
^{\varepsilon }\overline{H} =\sqrt{\left\vert \frac{\overline{\rho }}{3}(1-%
\frac{\overline{\rho }}{\overline{\rho }_{c}})\right\vert }.
\end{equation*}
}

The possibility to induce and preserve certain anamorphic cosmological
phases following the conditions (\ref{anamcr}). For mixed 3 type QC
structures, the Weyl type anamorphic $\varepsilon $--deformed
characteristics are determined also by the data for the integration function
$h_{4}^{[0]}(x^{k});$ effective sources $^{F}\mathbf{\Upsilon ,}~^{m}\mathbf{%
\Upsilon ,}~\overline{\mathbf{\Upsilon }}$ and $\overline{\rho }$ contained
in the sum $\ ^{\varepsilon }\widehat{H}+\ ^{\varepsilon }\overline{H}.$ We
compute {\small
\begin{equation*}
\begin{tabular}{lllll}
&  & $\mbox{ anamorphosis }$ & $\mbox{ inflation }$ & $\mbox{ ekpyrosis }$
\\
&  &  &  &  \\
$M_{Pl}^{0}~_{\varepsilon }^{m}\Theta =\frac{(\ln |\sqrt{|\ ^{\varepsilon
}h_{4}|}(1+\varepsilon \chi )|)^{\ast }}{(1+\varepsilon \chi _{PL})}$ &  & $%
<0\mbox{ (contracts)  }$ & $>0\mbox{
(expands) }$ & $<0\mbox{ (contracts)  }$ \\
$M_{Pl}^{0}~_{\varepsilon }^{Pl}\Theta =\frac{(\ln |\sqrt{|\ ^{\varepsilon
}h_{4}|}(1+\varepsilon \chi _{PL})|)^{\ast }}{(1+\varepsilon \chi _{PL})}$ &
& $>0\mbox{ (grow) }$ & $>0\mbox{
(grow) }$ & $>0\mbox{ (decay) }$%
\end{tabular}%
\end{equation*}%
} In such criteria, we use the value $\ ^{\varepsilon }h_{4}$ (\ref{h4e})
conditions imposing additional nonholonomic constraints on generating
functions and all types of sources and integration functions and constants
which induce QC structures. In a similar form, we can generate $\varepsilon $%
-analogs of (\ref{ex1}), (\ref{ex2}) and (\ref{ex3}), (\ref{enadcbc}) and
analyze if respective conditions for anamorphic phases can be satisfied.

In references \cite{bamba,amoros}, detailed studies using analytical methods
and numerical computations for holonomy corrections were performed in order
to demonstrate respectively how such terms may prevent the Big Rip
singularity in LQC and how to avoid singularities in certain MGTs. It was
concluded that the dynamics with holonomy corrections is very different from
that for original $R+\alpha R^{2}$ models (for such theories, "the universe
is singular at early times and never bounces"). The research in Refs. \cite%
{zhang1,zhang2,amoros1,amoros2,vlqgdq,bamba,amoros} was performed for
diagonal ansatz constraining from the very beginning the possibility to
generate quasi-periodic, pattern forming, and/or 3-d soliton cosmological
configurations depending, in principle, on all spacetime coordinates. The
AFDM (see a survey in Appendix \ref{appendixsect}, and references therein) allows us to
construct very general classes of locally anisotropic and inhomogeneous
cosmological solutions in varios types of MGTs \cite%
{cosmv2,eegrg,eeejpp,cosmv3,partn1,partn2,partn3}. Such generic off-diagonal
cosmological models may describe anamorphic phases with quasi-periodic structures, singular or nonsigular configurations etc. It was not clear if the results and conclusions on holonomy corrections, for
instance, those obtained in \cite{bamba,amoros} hold true for generic
off-diagonal configurations. The results of this section prove that at least
for off-diagonal $\varepsilon $--deformations the holonomy corrections may also prevent/
avoid singularities (as in the cases isotropic and homogeneous configurations).
Finally we note that the holonomy corrections  $\overline{H}%
(x^{i},t)=\sin (\sqrt{2\sqrt{3\gamma }}\overline{\beta }(x^{i},t))/\sqrt{2%
\sqrt{3\gamma }}$ considered section \ref{sseclqc} can be computed for arbitrary
classes of off-diagonal solutions. The conclusion on avoiding or generating
singularities depends on the type of nonlinear cosmological configurations
we define by generating functions and/or effective sources. This should be
analyzed in explicit form for a chosen example of cosmological metric, for instance,
determined by a cosmological solitonic configuration, or a pattern forming structure.

\section{Concluding Remarks}

\label{secconcl}

The Planck temperature anistoropy maps were used to probe the large-scale
spacetime structure \cite{planck15a26}. The observational data
were completed with respective calculus for the Baysesian likelihood with
simulations for specific topological models (in universes with locally flat,
hyperbolic and spherical geometries). All such work found no evidence for a
multiply--connected spacetime topology (when the assumption on the
fundamental domain is considered within the last scattering surface). No
matching circles, which would result from the intersection of fundamental
topological domains with the surface of last scattering, were found. It is
supposed that future Planck measurements of CMB polarization may provide
more definitive conclusions on anisotropic geometries and non-trivial
topologies. At present, the Planck data provides certain phenomenological
evidence for a Bianchi $VII_{h}$ component when parameters are decoupled
from standard cosmology. There is no a well defined set of cosmological
parameters which can produce existing patterns and observed anisotropies on
other scales.

Following new results of Planck2015 \cite%
{planck15a13,planck15a14,planck15a20,planck15a26,planck15a31} ( with the
ratio of tensor perturbation amplitude $r<0.1$) authors of \cite%
{sht1a,sht2a,sht3a,sht4a,sht5a,ijj} concluded that such observational data
seem to "virtually eliminate all the simplest textbook inflationary models".
In order to solve this problem and update cosmological scenarios, theorists
elaborate \cite{guthah01,kallosh1,mukh01,starob01} on three classes of
cosmological theories:

\begin{itemize}
\item There are alternative plateau-like and multi-parameter models adjusted
in such ways that necessary $r$ is reproduced. This results in new
challenges like 'unlikeness' and multiverse--un predictiabi\-lity problems with
more tuning and of parameters and initial conditions.

\item The classic inflationary paradigm is changed into a 'postmodern' one
and a MGT that allow certain flexibility to fit any combination of
observations. Even a series of conceptual problem of initial conditions and
multiverse is known and unresolved for decades, many theorists still
advocate this direction.

\item There are developed "bouncing" cosmologies, for instance, certain
versions of ekpyrotic (cyclic) cosmology and, also, anamorphic cosmology. In
such models, the large scale structure of the universe is set via a period
of slow contraction when the big bang is replaced by a big bounce. The
anamorphic approach is also considered as a different scenario with a
smoothing and flattening of the universe via a contracting phase. This way,
a nearly scale-ivariant spectrum of perturbations is generated.
\end{itemize}

The ekpyrotic cosmology \cite{sht4a} fits quite well the Planck2015 data
even in the simplest version with the least numbers of parameters and the
least amount of tuning. It provides a mechanism for getting a smooth and
flat cosmological background via a period of ultra slow contraction before
the big bang. For such a model, there are not required improbable initial
conditions and the multiverse problem is avoided. Realistic ekpyrotic
theories \cite{sht4a,sht4,sht5,sht6} involve two scalar fields when only one
has a negative potential in such a form that a non-canonical kinetic
coupling acts as an additional friction term for a scalar field freezing the
second one. A standard stability analysis proves that diagonal cosmological
solutions for such a model are scale-invariant and stable.

We note that the anamorphic cosmology \cite%
{sht1a,sht2a,sht3a,sht4a,sht5a,ijj} was developed as an attempt to describe
the early-universe in a form combining the advantages of the "old and
modern" inflationary and ekpyrotic models. The main assumption is that the
effective Plank mass, $M_{Pl}(t)$, has a different time dependence on $t$,
compared to the mass of a massive particle $m(t)$ in any Weyl frame during
the primordial genesis phase. Such cosmological models with similar, or
different, variations of fundamental constants and masses of particles can
be developed in the framework of various MGTs, see discussions in \cite%
{sht1a,sht3a}. In our works \cite{cosmv2,eegrg,eeejpp,cosmv3,vrajssrf}, we
proved that it is possible to construct exact solutions with effective
polarization of constants (in general, depending on all spacetime
coordinates, $M_{Pl}(x^{i},y^{3},t)$ and $m(x^{i},y^{3},t)$) in GR mimicking
time-like dependencies in MGTs if generic off--diagonal metrics and
nonholonomically deformations of connections are considered for constructing
new classes of cosmological solutions. Such exact/ parametric solutions can
be constructed in general form using the anholonomic frame deformation
method,\ AFDM, see review of results in \cite{gvvafdm,vanp} and references
therein. Following this geometric method, we perform such nonholonomic
deformations of the coefficients of frames, generic off-diagonal metrics and
(generalized) connections when the (generalized) Einstein equations can be
decoupled in general forms and integrated for various classes of metrics $%
g_{\alpha \beta }(x^{i},y^{3},t).$

Finally, we note that noholonomic anamorphic scenarios allow us to preserve
the paradigm of Einstein's GR theory and to produce cosmological (expanding
for certain phases and contracting in other cases) inflation and
acceleration, if generic off-diagonal gravitational interactions model
equivalently modifications of diagonal configurations in MGTs. This is
possible if more general classes of cosmological solutions encoding QC
structures and LQG corrections are considered.

\vskip5pt \textbf{Acknowledgments:}\ SV research on geometric methods for constructing exact solutions in gravity theories and applications in modern cosmology and astrophysics was supported by programs IDEI,
PN-II-ID-PCE-2011-3-0256, and former senior research fellowships from CERN, and DAAD. In a particular case, certain examples of  quasi-periodic and aperiodic cosmological solutions were elaborated following  a
Consulting Agreement (for a visiting researcher) with a private organization QGR-Topanga, California, the USA, during May 18, 2016 - June 2, 2017.

\appendix

\setcounter{equation}{0} \renewcommand{\theequation}
{A.\arabic{equation}} \setcounter{subsection}{0}
\renewcommand{\thesubsection}
{A.\arabic{subsection}}

\section{Off-diagonal Cosmological Solutions in MGTs}

\label{appendixsect} We present a brief review of the anholonomic frame
deformation method, AFDM, for generating off-diagonal solutions in MGTs and
GR, see details  in \cite{cosmv3,gvvafdm,eeejpp,eegrg} and
references therein.

The N-adapted coefficients of the canonical d-connection $\widehat{\mathbf{D}%
}=\{\widehat{\mathbf{\Gamma }}_{\ \alpha \beta }^{\gamma }=(\widehat{L}%
_{jk}^{i},\widehat{L}_{bk}^{a},\widehat{C}_{jc}^{i},\widehat{C}_{bc}^{a})\}$
(\ref{twocon}) are computed following formulas
\begin{eqnarray}
\widehat{L}_{jk}^{i} &=&\frac{1}{2}g^{ir}\left( \mathbf{e}_{k}g_{jr}+\mathbf{%
e}_{j}g_{kr}-\mathbf{e}_{r}g_{jk}\right) ,\widehat{C}_{bc}^{a}=\frac{1}{2}%
g^{ad}\left( e_{c}g_{bd}+e_{b}g_{cd}-e_{d}g_{bc}\right) ,  \label{candcon} \\
\widehat{C}_{jc}^{i} &=&\frac{1}{2}g^{ik}e_{c}g_{jk},\ \widehat{L}%
_{bk}^{a}=e_{b}(N_{k}^{a})+\frac{1}{2}g^{ac}\left( \mathbf{e}%
_{k}g_{bc}-g_{dc}\ e_{b}N_{k}^{d}-g_{db}\ e_{c}N_{k}^{d}\right) .  \notag
\end{eqnarray}%

The torsion, $\widehat{\mathcal{T}},$ and the curvature, $\widehat{\mathcal{R%
}}$, tensors of $\widehat{\mathbf{D}}=(h\widehat{D},v\widehat{D})$ are
defined in standard froms for any distinguished vectors, d-vectors, $\mathbf{%
X}$ and $\mathbf{Y,}$
\begin{equation*}
\widehat{\mathcal{T}}(\mathbf{X,Y}):=\widehat{\mathbf{D}}_{\mathbf{X}}%
\mathbf{Y}-\widehat{\mathbf{D}}_{\mathbf{Y}}\mathbf{X}-[\mathbf{X,Y}]%
\mbox{
and }\mathcal{R}(\mathbf{X,Y}):=\widehat{\mathbf{D}}_{\mathbf{X}}\widehat{%
\mathbf{D}}_{\mathbf{Y}}-\widehat{\mathbf{D}}_{\mathbf{Y}}\widehat{\mathbf{D}%
}_{\mathbf{X}}-\widehat{\mathbf{D}}_{\mathbf{[X,Y]}}.
\end{equation*}%
Such formulas (including definitions of the Ricci d-tensor and related
scalar curvature) can be written and computed in N-adapted form as in
footnote \ref{notericci}. For (\ref{candcon}), we express the nontrivial
d--torsion coefficients\ $\widehat{\mathbf{T}}_{\ \alpha \beta }^{\gamma }$
in the form
 $\widehat{T}_{\ jk}^{i}=\widehat{L}_{jk}^{i}-\widehat{L}_{kj}^{i},\,\,\,%
\widehat{T}_{\ ja}^{i}=\widehat{C}_{jb}^{i},\,\,\,\widehat{T}_{\
ji}^{a}=-\Omega _{\ ji}^{a},\,\,\,\widehat{T}_{aj}^{c}=\widehat{L}%
_{aj}^{c}-e_{a}(N_{j}^{c}),\,\,\,\widehat{T}_{\ bc}^{a}=\ \widehat{C}%
_{bc}^{a}-\ \widehat{C}_{cb}^{a}$.
  These d-torsion coefficients vanish if there are
satisfied the conditions
\begin{equation}
\widehat{L}_{aj}^{c}=e_{a}(N_{j}^{c}),\,\,\,\widehat{C}_{jb}^{i}=0,\Omega
_{\ ji}^{a}=0.  \label{lccond}
\end{equation}%
Using above formulas for (\ref{candcon}) and any d-metric (\ref{odans}), we
can compute the coefficients of the Riemann d-tensor, $\widehat{\mathbf{R}}%
_{\ \beta \gamma \delta }^{\alpha },$ the Ricci d-tensor, $\widehat{\mathbf{R%
}}_{\alpha \beta }$, and the Einstein d-tensor $\widehat{\mathbf{E}}_{\alpha
\beta }:=\widehat{\mathbf{R}}_{\alpha \beta }-\frac{1}{2}\mathbf{g}_{\alpha
\beta }\ \widehat{R}.$ Such values can be similarly computed for a
LC-connection $\nabla =\{\Gamma _{\ \alpha \beta }^{\gamma }\}$.

To generate locally anisotropic and nonhomogeneous cosmological solutions, we
consider a d--metric $\mathbf{g}$ (\ref{odans}) which via frame and
coordinate transforms can be parameterized in the form
\begin{equation}
g_{i}=e^{\psi {(x^{k})}},\,\,\,\,g_{a}=\omega (x^{k},y^{b})h_{a}(x^{k},t),\
N_{i}^{3}=n_{i}(x^{k},t),\,\,\,\,N_{i}^{4}=w_{i}(x^{k},t).  \label{data1c}
\end{equation}%
 For $\omega =1,$ such off-diagonal cosmological metrics posses a space
like killing symmetry on $\partial _{3}=\partial q/\partial \varphi .$ We
can use brief notations of partial derivatives $\partial _{\alpha
}q=\partial q/\partial u^{\alpha }$ when, for any function $q(x^{k},y^{a}),$
one compute $\partial _{1}q=q^{\bullet }=\partial q/\partial x^{1},\partial
_{2}q=q^{\prime }=\partial q/\partial x^{2},\partial _{3}q=\partial
q/\partial y^{3}=\partial q/\partial \varphi =q^{\diamond },\partial
_{4}q=\partial q/\partial t=\partial _{t}q=q^{\ast },\partial
_{33}^{2}=\partial ^{2}q/\partial \varphi ^{2}=\partial _{\varphi \varphi
}^{2}q=q^{\diamond \diamond },\partial _{44}^{2}=\partial ^{2}q/\partial
t^{2}=\partial _{tt}^{2}q=q^{\ast \ast }.$ The sources (\ref{sourcc}) for
(effective) matter field configurations can be parameterized via frame
transforms in respective N--adapted forms, $\mathbf{\Upsilon }_{\ \nu }^{\mu
}=\mathbf{e}_{\ \mu ^{\prime }}^{\mu }\mathbf{e}_{\nu }^{\ \nu ^{\prime }}~%
\mathbf{\Upsilon }_{\ \nu ^{\prime }}^{\mu ^{\prime }}]=[~\ _{h}\Upsilon
(x^{i})\delta _{j}^{i},\Upsilon (x^{i},t)\delta _{b}^{a}]$ (\ref{sources}).
The values $~\ _{h}\Upsilon (x^{i})$ and $\Upsilon (x^{i},t)]$ can be taken
as generating functions for (effective) matter sources. They impose
nonholonomic frame constraints on cosmological dynamics of (effective)
matter fields. For simplicity, we consider generation of generic
off--diagonal cosmological solutions with $\partial _{b}h_{a}\neq 0$ and $%
[~\ _{h}\overline{\Upsilon }(x^{i}),\overline{\Upsilon }(x^{i},t)]\neq 0$
(such conditions can be satisfied for certain frame/coordianate systems).%
\footnote{It is possible to construct important (non) vacuum solutions if such
conditions are not satisfied with respect to certain systems of references.
This requests more special methods.}

Let us prove that above introduced parameterizations of cosmological
d--metrics and (effective) sources allows us to integrate in explicit form
the MGT field equations (\ref{mfeq}). Here we note that constructing
off-diagonal solutions for some nonholonomic cosmological configurations we
can impose additional nonholonomic constraints and obtain configurations
with $g_{\alpha \beta }(x^{i},t)\approx g_{\alpha \beta }(t)$ which can be
related to Bianchi type, or FLRW, like cosmological metrics.

Considering d-metrics with data (\ref{data1c}) for $\omega =1,$ we compute $%
\widehat{\mathbf{D}}=\{\widehat{\mathbf{\Gamma }}_{\ \alpha \beta }^{\gamma
}\}$ (\ref{candcon}) and $\widehat{\mathbf{R}}_{\alpha \beta }.$ The
modified Einstein equations (\ref{mfeq}) transform into a system of
nonlinear PDEs,
\begin{eqnarray}
\widehat{R}_{1}^{1} &=&\widehat{R}_{2}^{2}=\frac{1}{2g_{1}g_{2}}[\frac{%
g_{1}^{\bullet }g_{2}^{\bullet }}{2g_{1}}+\frac{\left( g_{2}^{\bullet
}\right) ^{2}}{2g_{2}}-g_{2}^{\bullet \bullet }+\frac{g_{1}^{\prime
}g_{2}^{\prime }}{2g_{2}}+\frac{(g_{1}^{\prime })^{2}}{2g_{1}}-g_{1}^{\prime
\prime }]=-\ ~\ _{h}\Upsilon ,  \label{eq1b} \\
\widehat{R}_{3}^{3} &=&\widehat{R}_{4}^{4}=\frac{1}{2h_{3}h_{4}}[\frac{%
\left( h_{3}^{\ast }\right) ^{2}}{2h_{3}}+\frac{h_{3}^{\ast }h_{4}^{\ast }}{%
2h_{4}}-h_{3}^{\ast \ast }]=-\Upsilon   \notag \\
\widehat{R}_{3k} &=&\frac{h_{3}}{2h_{4}}n_{k}^{\ast \ast }+(\frac{h_{3}}{%
h_{4}}h_{4}^{\ast }-\frac{3}{2}h_{3}^{\ast })\frac{n_{k}^{\ast }}{2h_{4}}=0,
  \notag \\
\widehat{R}_{4k} &=&-\frac{w_{k}}{2h_{3}}[\frac{\left( h_{3}^{\ast }\right)
^{2}}{2h_{3}}+\frac{h_{3}^{\ast }h_{4}^{\ast }}{2h_{4}}-h_{3}^{\ast \ast }]+%
\frac{h_{3}^{\ast }}{4h_{3}}(\frac{\partial _{k}h_{3}}{h_{3}}+\frac{\partial
_{k}h_{4}}{h_{4}})-\frac{\partial _{k}h_{3}^{\ast }}{2h_{3}}=0,   \notag
\end{eqnarray}%
for partial derivatives $\partial _{t}q=$ $\partial _{4}q=q^{\ast }$ and $%
\partial _{i}q=(\partial _{1}q=q^{\bullet },\partial _{2}q=q^{\prime }).$
The zero torsion conditions request impose for data (\ref{data1c}) additional
LC--conditions (\ref{lccond}) which can be written in the form
\begin{equation}
\partial _{t}w_{i}=(\partial _{i}-w_{i}\partial _{t})\ln \sqrt{|h_{4}|}%
,(\partial _{i}-w_{i}\partial _{t})\ln \sqrt{|h_{3}|}=0,\partial
_{k}w_{i}=\partial _{i}w_{k},\partial _{t}n_{i}=0,\partial
_{i}n_{k}=\partial _{k}n_{i}.  \label{lccondb}
\end{equation}

The system (\ref{eq1b}) can be written in the form
\begin{eqnarray}
\psi ^{\bullet \bullet }+\psi ^{\prime \prime } &=&2\ ~\ _{h}\Upsilon
\label{e1} \\
{\varpi }^{\ast }h_{3}^{\ast } &=&2h_{3}h_{4}\Upsilon    \notag \\
n_{i}^{\ast \ast }+\gamma n_{i}^{\ast } &=&0,   \notag \\
\beta w_{i}-\alpha _{i} &=&0,   \notag
\end{eqnarray}%
where the system of reference is chosen for $\partial _{t}h_{a}\neq 0$ and $%
\partial _{t}\varpi \neq 0$ and the coefficients are computed,
\begin{equation}
\alpha _{i} = (\partial _{t}h_{3})\ (\partial _{i}\varpi ),\ \beta
=(\partial _{t}h_{3})\ (\partial _{t}\varpi ),\ \gamma =\partial _{t}\left(
\ln |h_{3}|^{3/2}/|h_{4}|\right) ,  \mbox{ where }\varpi ={\ln |\partial _{t}h_{3}/\sqrt{|h_{3}h_{4}|}|}.
\label{genf1b}
\end{equation}%
The system of nonlinear PDEs (\ref{e1}) reflects a decoupling property of equations for functions $\psi ,h_{a},n_{i}^{\ast }$ and can be integrated in general form  for any generating function $\Psi
(x^{i},t):=e^{\varpi }$ and sources $\ _{h}\Upsilon (x^{i})$ and $\Upsilon (x^{k},t).$

The system (\ref{e1}) can be integrating in general form "step by step". In result, we generate exact solutions of the modified Einstein equations (\ref{mfeq}) parameterized by such coefficients of a
\begin{eqnarray}
\mbox{d--metric: } \ g_{i} &=&e^{\ \psi (x^{k})}\mbox{ as a solution of 2-d Poisson eqs. }\psi
^{\bullet \bullet }+\psi ^{\prime \prime }=2~\ _{h}\Upsilon ;  \notag \\
g_{3} &=&h_{3}({x}^{i},t)=-(\Psi ^{2})^{\ast }/4\Upsilon ^{2}h_{4}=-(\Psi
^{2})^{\ast }/4\Upsilon ^{2}(h_{4}^{[0]}(x^{k})-\int dt(\Psi ^{2})^{\ast
}/4\Upsilon )  \label{offdcosm} \\
g_{4} &=&h_{4}({x}^{i},t)=h_{4}^{[0]}(x^{k})-\int dt(\Psi ^{2})^{\ast
}/4\Upsilon ;  \notag \\
\mbox{N--connection: }
 N_{k}^{3} &=&n_{k}({x}^{i},t)=\ _{1}n_{k}(x^{i})+\ _{2}n_{k}(x^{i})\int
dt(\Psi ^{\ast })^{2}/\Upsilon ^{2}|h_{4}^{[0]}(x^{i})-\int dt(\Psi
^{2})^{\ast }/4\Upsilon |^{5/2};  \notag \\
N_{i}^{4} &=&w_{i}({x}^{i},t)=\partial _{i}\Psi /\Psi ^{\ast }=\partial
_{i}\Psi ^{2}/\ (\Psi ^{2})^{\ast }.  \notag
\end{eqnarray}%
In these formulas, the values $h_{4}^{[0]}(x^{k}),$ $\ _{1}n_{k}(x^{i}),$
and $\ _{2}n_{k}(x^{i})$ are integration functions. The
coefficients (\ref{offdcosm}) define generic off-diagonal cosmological
solutions if some anholonomy coefficients $C_{\alpha \beta }^{\gamma
}(x^{i},t)$ (\ref{anhr}) are not zero. Such locally cosmological solutions
can be with nontrivial nonholonomically induced d-torsion or for
LC-configurations if the conditions (\ref{lccondb}) are satisfied. We
generate as particular cases some well-known cosmological FLRW, or Bianchi,
type metrics, for cerain  data of type $(\Psi (t),\Upsilon (t))$ with
integration functions which allow frame/ coordinate transforms to respective
(off-) diagonal configurations $g_{\alpha \beta }(t).$

Introducing the coefficients (\ref{offdcosm}) into (\ref{data1c}) $\mathbf{g}
$ (\ref{odans}), we construct construct a class of linear quadratic elements
for  locally anisotropic cosmological solutions,
\begin{eqnarray}
ds^{2} &=&e^{\ \psi (x^{k})}[(dx^{1})^{2}+(dx^{2})^{2}]
+(h_{4}^{[0]}-\int dt\frac{(\overline{\Psi }^{2})^{\ast }}{4\overline{%
\Upsilon }})[dt+\frac{\partial _{i}\ \overline{\Psi }}{\ \overline{\Psi }%
^{\ast }}dx^{i}]  \label{gensolcos} \\
&& -\frac{(\overline{\Psi }^{2})^{\ast }}{4\Upsilon ^{2}\left( h_{4}^{[0]}-\int
dt\frac{(\overline{\Psi }^{2})^{\ast }}{4\overline{\Upsilon }}\right) }%
[dy^{3}+(_{1}n_{k}+\ _{2}n_{k}\int dt\frac{(\overline{\Psi }^{\ast })^{2}}{%
\overline{\Upsilon }^{2}|h_{4}^{[0]}-\int dt\frac{(\overline{\Psi }%
^{2})^{\ast }}{4\overline{\Upsilon }}|^{5/2}})dx^{k}]. \notag
\end{eqnarray}
Such solutions posses a Killing symmetry on $\partial _{3}$ and can be
re-written in terms of $\eta $--polarization function functions for target
locally anisotropic cosmological metrics $\ \widehat{\mathbf{g}}\mathbf{=}%
[g_{\alpha }=\eta _{\alpha }\mathring{g}_{\alpha },\ \eta _{i}^{a}\mathring{N%
}_{i}^{a}]$ encoding primary cosmological data $[\mathring{g}_{\alpha },%
\mathring{N}_{i}^{a}].$

We can extract cosmological spacetimes in GR (with zero torsion) if the
conditions (\ref{lccondb}) are imposed and solved for a special class of
generating functions and sources. For instance, taking a $\Psi =\check{\Psi}%
(x^{i},t)$ subjected to the conditions $(\partial _{i}\check{\Psi})^{\ast
}=\partial _{i}(\check{\Psi}^{\ast })$ and $\Upsilon (x^{i},t)=\Upsilon
\lbrack \check{\Psi}]=\check{\Upsilon},$ or $\Upsilon =const,$ we generate
LC--configurations for some functions $\check{A}(x^{i},t)$ and $n(x^{i})$
when the N--connection coefficients are computed $\ \overline{n}_{k}=\check{n%
}_{k}=\partial _{k}n(x^{i})$ and $w_{i}=\partial _{i}\check{A}=\partial _{i}%
\check{\Psi}/\check{\Psi}^{\ast }.$ Such off-diagonal locally anisotropic
cosmological solutions in GR are defined as subclasses of solutions (\ref%
{gensolcos}) with zero torsion,%
\begin{equation}
ds^{2}=e^{\ \psi (x^{k})}[(dx^{1})^{2}+(dx^{2})^{2}]-\frac{(\check{\Psi}%
^{2})^{\ast }}{4\check{\Upsilon}^{2}(h_{4}^{[0]}-\int dt\frac{(\check{\Psi}%
^{2})^{\ast }}{4\check{\Upsilon}})}[dy^{3}+(\partial
_{k}n)dx^{k}]+(h_{4}^{[0]}-\int dt\frac{(\check{\Psi}^{2})^{\ast }}{4\check{%
\Upsilon}})[dt+(\partial _{i}\check{A})dx^{i}].  \label{lcsolcos}
\end{equation}

Quadratic linear elements for exact off-diagonal solutions (\ref{gensolcos})
or (\ref{lcsolcos}) constructed above can be parameterized in
the form (\ref{targm}). Such parameterizations  are in terms of polarization
functions $\eta _{\alpha }=(\eta _{i},\eta _{a})$ and $\eta _{i}^{a}$
defining nonholonomic deformations of a prime d-metric, $\mathbf{\mathring{g}%
,}$ into a target d-metric, $\widehat{\mathbf{g}}=[g_{\alpha }=\eta _{\alpha
}\mathring{g}_{\alpha },\ \eta _{i}^{a}\mathring{N}_{i}^{a}]\rightarrow
\mathbf{\mathring{g}}$. Such parameterizations with $\varepsilon $%
-deformations are useful for analyzing possible physical implications of
general off-diagonal deformations of some physically important solutions
when, for instance, $\mathbf{\mathring{g}}$ is taken for a standard
cosmological, solution in GR, or in a MGT.

Finally, we note that the AFDM allows to construct off-diagonal cosmological solutions in  general form depending in arbitrary classes of generating functions and effective sources.  Such functions may remove existing singularities of a prime metric, or inversely, to transform nonsingular configurations into singular locally anisotropic cosmological ones because of singular nonholonomic deformations. In general, it is not clear what physical implications may have some classes off-diagonal solutions. In section  \ref{sectodancosm}, we demonstrate that  well-defined physical interpretations can be provided for anamorphic cosmological  quasi-periodic, pattern-forming and/or solitonic structures generated in in general off-diagonal form.  For $\epsilon$-deformations studied in sections \ref{sspard} and  \ref{epsilonpar}, such locally anisotropic cosmological structures can be considered for modeling dark matter and dark energy effects  determined by "small" off-diagonal deformations of FLRW, or Bianchi, type metrics in standard cosmology.

\end{document}